\documentclass[12pt]{article}
\usepackage{amsfonts}
\usepackage{graphicx}
\usepackage{amsmath}
\usepackage{float}
\usepackage{amssymb}
\providecommand{\U}[1]{\protect\rule{.1in}{.1in}}
\textwidth=6.5in \hoffset=-.75in \voffset=-1in \numberwithin{equation}{section}
\numberwithin{figure}{section}

\begin{document}
\begin{titlepage}
\bigskip \begin{flushright}
\end{flushright}
\vspace{1cm}
\begin{center}
{\Large \bf {Gibbons-Hawking M-branes}}\\
\end{center}
\vspace{1cm}
\begin{center}
A. M.
Ghezelbash{ \footnote{ E-Mail: masoud.ghezelbash@usask.ca}}, R. Oraji
{ \footnote{ E-Mail: rao519@mail.usask.ca}}
\\
Department of Physics and Engineering Physics, \\ University of Saskatchewan,
Saskatoon, Saskatchewan S7N 5E2, Canada\\
\vspace{1cm}
\end{center}
\begin{abstract}

We present new M2 and M5-brane solutions in M-theory based on transverse 
Gibbons-Hawking spaces. These solutions provide realizations of fully localized type IIA D2/D6 and NS5/D6 brane
intersections. One novel feature of these 
solutions is that the metric functions depend on more than two transverse coordinates (unlike all the other previous known solutions). All the solutions have eight preserved supersymmetries and the
world-volume theories of the NS5-branes are new non-local, non-gravitational, six dimensional, T-dual little
string theories with eight supersymmetries. We discuss the limits in which the
dynamics of the D2 and NS5-branes decouple from the bulk for these solutions.

\end{abstract}
\bigskip
\end{titlepage}\onecolumn

\bigskip

\section{Introduction}

Fundamental M-theory in the low-energy limit is generally believed to be
effectively described by $D=11$ supergravity \cite{gr1,gr2,gr3}. This suggests
that brane solutions in the latter theory furnish classical soliton states of
M-theory, motivating considerable interest in this subject. There is
particular interest in finding $D=11$ M-brane solutions that reduce to
supersymmetric $p$-brane solutions (that saturate the
Bogomol'nyi-Prasad-Sommerfield (BPS) bound) upon reduction to 10 dimensions.
Some supersymmetric BPS solutions of two or three orthogonally intersecting
2-branes and 5-branes in $D=11$\ supergravity were obtained some years ago
\cite{Tsey}, and more such solutions have since been found \cite{oth}.

Recently interesting new supergravity solutions for localized D2/D6, D2/D4,
NS5/D6 and NS5/D5 intersecting brane systems were obtained
\cite{hashi,CGMM2,ATM2,GHEbianchiMbranes,GH2resolvedconifolds}. By lifting a
D6 (D5 or D4)-brane to four-dimensional self-dual geometries embedded in
M-theory, these solutions were constructed by placing M2- and M5-branes in
different self-dual geometries. A special feature of this construction\ is
that the solution is not restricted to be in the near core region of the D6
(or D5) brane, a feature quite distinct from the previously known solutions
\cite{IT,IT2}. For all of the different BPS solutions, 1/4 of the supersymmetry is
preserved as a result of the self-duality of the transverse metric. Moreover,
in \cite{Ali}, partially localized D-brane systems involving D3, D4 and D5
branes were constructed. By assuming a simple ansatz for the eleven
dimensional metric, the problem reduces to a partial differential equation
that is separable and admits proper boundary conditions.

Motivated by this work, the aim of this paper is to construct the fully
localized supergravity solutions of D2 (and NS5) intersecting D6 branes without restricting to
the near core region of the D6 by reduction of ALE geometries lifted to
M-theory. Our main motivation for considering ALE geometries (and specially
multi-center Gibbons-Hawking spaces) is that in all previously constructed
M-brane solutions
\cite{hashi,CGMM2,ATM2,GHEbianchiMbranes,GH2resolvedconifolds}, we have at
most one parameter in each solution. For example, NUT/Bolt parameter $n$ for
embedded transverse Taub-NUT/Bolt spaces, Eguchi-Hanson parameter $a$ in the
case of embedded transverse Eguchi-Hanson geometry and a constant number with
unit of length that is related to the NUT charge of metric at infinity
obtained from Atiyah-Hitchin metric in the case of embedded transverse
Atiyah-Hitchin geometry. Moreover, in all the above mentioned solutions, the
metric functions depend (at most) only on two non-compact coordinates. The
metric functions in the multi-center Gibbons-Hawking geometries depend (in
general) on more physical parameters, hence their embeddings into M-theory
yield new results for the metric functions with both non-compact and compact coordinates.

We have obtained several different supersymmetric BPS solutions of interest. We should mention the condition of preserved supersymmetry is distinct from that of BPS which is defined in the bosonic theory. Due to the general M2 and M5 ansatze that we consider in sections \ref{sec:MbraneGH},\ref{sec:M5} and \ref{sec:M2mix}, the metric functions for all M2 solutions, as well as M5 solutions are harmonic. Hence all our brane configurations are determined by solutions of Laplace equations and so they obey the BPS property. Specifically, since
in the 11 dimensional metric for an M2-brane, the M2-brane itself only takes
up two of the 10 spatial coordinates, we can embed a variety of different
geometries. These include the double Taub-NUT metric, two-center
Eguchi-Hanson metric and products of these 4-dimensional metrics. After
compactification on a circle, we find the different fields of type IIA string theory.

In our procedure we begin with a general ansatz for the metric function of an
M2 brane in 11-dimensional M-theory. After compactification on a circle
$(T^{1}),$ we find a solution to type IIA theory for which the highest degree
of the field strengths is four. Hence the non-compact global symmetry for
massless modes is given by the maximal symmetry group $E_{1(1)}=\mathbb{R}$,
without any need to dualize the field strengths \cite{cremmer}. For the full
type IIA theory, only the discrete subgroup $E_{1(1)}(\mathbb{Z})=\mathbb{Z}$
survives, in particular by its action on the BPS spectrum and as a discrete
set of identifications on the supergravity moduli space. This subgroup is the
U-duality group for all type IIA theories we find in this paper.{\large \ }

The outline of our paper is as follows. In section \ref{sec:review}, we discuss briefly the field
equations of supergravity. In section \ref{sec:GH} we review briefly the ALE
geometries and then in section \ref{sec:MbraneGH}, we consider the embedding
of four-dimensional multi (and explicitly double) -center Gibbons-Hawking spaces in
M-theory. These spaces are characterized with some
(two) NUT charges. Moreover, we consider the multi (and especially two-center)
Eguchi-Hanson spaces and find analytical exact solutions for the M2-brane functions. We compare then our analytical solutions with the numerical solutions found a few years ago. 
In section \ref{sec:M5}, we present the M5-brane solutions. These solutions also are exact and analytic. 
In section \ref{sec:M2mix}, we then discuss embedding products of Gibbons-Hawking metrics in
M2-brane solutions. All of the solutions preserve some of the
supersymmetry as we present the details in section \ref{sec:susy}. In section \ref{sec:dec}, we consider the decoupling limit of
our solutions and find evidence that 
in the limit of vanishing string
coupling, the theory on the world-volume of the NS5-branes is a new little
string theory. Moreover, we apply T-duality
transformations on type IIA solutions and find type IIB NS5/D5 intersecting
brane solutions and discuss the decoupling limit of the solutions.
We wrap up then by some concluding remarks and future possible research directions.

\section{Supergravity Solutions}

\label{sec:review} The equations of motion for eleven dimensional supergravity
when we have maximal symmetry (i.e. for which the expectation values of the
fermion fields is zero), are \cite{DuffKK}
\begin{align}
R_{mn}-\frac{1}{2}g_{mn}R  &  =\frac{1}{3}\left[  F_{mpqr}F_{n}
^{\phantom{n}pqr}-\frac{1}{8}g_{mn}F_{pqrs}F^{pqrs}\right] \label{GminGG}\\
\nabla_{m}F^{mnpq}  &  =-\frac{1}{576}\varepsilon^{m_{1}\ldots m_{8}
npq}F_{m_{1}\ldots m_{4}}F_{m_{5}\ldots m_{8}} \label{dF}
\end{align}
where the indices $m,n,\ldots$ are 11-dimensional world space indices. For an
M2-brane, we use the metric and four-form field strength
\begin{equation}
ds_{11}^{2}=H(y,r,\theta)^{-2/3}\left(  -dt^{2}+dx_{1}^{2}+dx_{2}^{2}\right)
+H(y,r,\theta)^{1/3}\left(  d\mathfrak{s}_{4}^{2}(y)+ds_{4}^{2}(r,\theta
)\right)  \label{ds11genM2}
\end{equation}
and non-vanishing four-form field components 
\begin{align}
F_{tx_{1}x_{2}y}  &  =-\frac{1}{2H^{2}}\frac{\partial H}{\partial y} 
\label{Fy}\\
F_{tx_{1}x_{2}r}  &  =-\frac{1}{2H^{2}}\frac{\partial H}{\partial r} 
\label{Fr}\\
F_{tx_{1}x_{2}\theta}  &  =-\frac{1}{2H^{2}}\frac{\partial H}{\partial\theta}
\label{Ft} 
\end{align}

and for an M5-brane, the metric and four-form field strength are
\begin{align}
ds^{2}  &  =H(y,r,\theta)^{-1/3}\left(  -dt^{2}+dx_{1}^{2}+\ldots+dx_{5} 
^{2}\right)  +H(y,r)^{2/3}\left(  dy^{2}+ds_{4}^{2}(r,\theta)\right)
~~~\label{ds11general}\\
F_{m_{1}\ldots m_{4}}  &  =\frac{\alpha}{2}\epsilon_{m_{1}\ldots m_{5} 
}\partial^{m_{5}}H~\ ~,~~~\alpha=\pm1 \label{Fgeneral} 
\end{align}
where $d\mathfrak{s}_{4}^{2}(y)$ and $ds_{4}^{2}(r)$ are two four-dimensional
(Euclideanized) metrics, depending on the non-compact coordinates $y$ and $r$,
respectively and the quantity $\alpha=\pm1,$\ which corresponds to an M5-brane
and an anti-M5-brane respectively. The general solution, where the transverse
coordinates are given by a flat metric, admits a solution with 16 Killing
spinors \cite{smith}.

The 11D metric and four-form field strength can be easily reduced down to ten
dimensions using the following equations 
\begin{align}
g_{mn}  &  =\left[
\begin{array}
[c]{cc} 
e^{-2\Phi/3}\left(  g_{\alpha\beta}+e^{2\Phi}C_{\alpha}C_{\beta}\right)  & \nu
e^{4\Phi/3}C_{\alpha}\\
\nu e^{4\Phi/3}C_{\beta} & \nu^{2}e^{4\Phi/3} 
\end{array}
\right] \label{FKKreduced1}\\
F_{(4)}  &  =\mathcal{F}_{(4)}+\mathcal{H}_{(3)}\wedge dx_{10}.
\label{FKKreduced2} 
\end{align}

Here $\nu$ is the winding number (the number of times the M-brane wraps
around the compactified dimensions) and $x_{10}$ is the eleventh dimension, on
which we compactify. The indices $\alpha,\beta,\cdots$ refer to ten-dimensional space-time components after compactification. $\mathcal{F}_{(4)}$ and $\mathcal{H}_{(3)}$ are the RR
four-form and the NSNS three-form field strengths corresponding to
$A_{\alpha\beta\gamma}$ and $B_{\alpha\beta}$.

The number of non-trivial solutions to the Killing spinor equation
\begin{equation}
\partial_{M}\varepsilon+\frac{1}{4}\omega_{abM}\Gamma^{ab}\varepsilon+\frac{1} 
{144}\Gamma_{M}^{\phantom{m}npqr}F_{npqr}\varepsilon-\frac{1}{18}\Gamma
^{pqr}F_{mpqr}\varepsilon=0 \label{killingspinoreq} 
\end{equation}
determine the amount of supersymmetry of the solution, where the $\omega$'s
are the spin connection coefficients, and $\Gamma^{a_{1}\ldots a_{n}} 
=\Gamma^{\lbrack a_{1}}\ldots\Gamma^{a_{n}]}$. The indices $a,b,...$ are 11
dimensional tangent space indices and the $\Gamma^{a}$ matrices are the eleven
dimensional equivalents of the four dimensional Dirac gamma matrices, and must
satisfy the Clifford algebra
\begin{equation}
\left\{  \Gamma^{a},\Gamma^{b}\right\}  =-2\eta^{ab}. \label{cliffalg} 
\end{equation}

In ten dimensional type IIA string theory, we can have D-branes or NS-branes.
D$p$-branes can carry either electric or magnetic charge with respect to the
RR fields; the metric takes the form 
\begin{equation}
ds_{10}^{2}=f^{-1/2}\left(  -dt^{2}+dx_{1}^{2}+\ldots+dx_{p}^{2}\right)
+f^{1/2}\left(  dx_{p+1}^{2}+\ldots+dx_{9}^{2}\right)  \label{gDpbrane} 
\end{equation}
where the harmonic function\thinspace\ $f$ generally depends on the transverse coordinates.

An NS5-brane carries a magnetic two-form charge; the corresponding metric has
the form
\begin{equation}
ds_{10}^{2}=-dt^{2}+dx_{1}^{2}+\ldots+dx_{5}^{2}+f\left(  dx_{6}^{2} 
+\ldots+dx_{9}^{2}\right).  \label{gNS5brane} 
\end{equation}
In what follows we will obtain a mixture of D-branes and NS-branes.

\section{Gibbons-Hawking Spaces}

\label{sec:GH}

The only instantons (in A-D-E classification) that their metrics could be
written in known closed forms, are $A_{k}$ series where the metrics are given
by:
\begin{equation}
ds^{2}=V^{-1}(dt+\vec A \cdot d \vec x)^{2}+V \gamma_{ij} dx^{i} \cdot d x^{j}
\label{Aseries} 
\end{equation}
where $V$, $A_{i}$ and $\gamma_{ij}$ are independent of $t$ and $\nabla
V=\pm\nabla\times\vec A $; hence $\nabla^{2} V=0$. The most general solution
for $V$ is then $V=\sum_{i=1}^{k} \frac{m}{\mid\vec x- \vec x _{i} \mid}$. The
metrics (\ref{Aseries}) describe the Gibbons-Hawking multi-center metrics. The
$k=1$ corresponds to flat space and $k=2$ corresponds to Eguchi-Hanson metric.
The standard form of Eguchi-Hanson metric is given by \cite{EgHan}
\begin{equation}
ds_{EH}^{2} =\frac{r^{2}}{4g(r)}\left[  d\psi+\cos(\theta)d\phi\right]
^{2}+g(r)dr^{2}+\frac{r^{2}}{4}\left(  d\theta^{2}+\sin^{2}(\theta)d\phi
^{2}\right)   \label{EHstandard} 
\end{equation}
where $g(r)=\frac{r^{4}}{r^{4}-a^{4}}$. If we change the coordinates of
(\ref{EHstandard}) to $(R,\Theta,\Phi,\Psi)$ by
\begin{align}
R  &  =\frac{1}{a}\sqrt{r^{4}-a^{4}\sin^{2} \theta}\label{C1}\\
\Theta &  =\tan^{-1}\big (\frac{\sqrt{r^{4}-a^{4}}}{r^{2}}\tan\theta
\big )\label{C2}\\
\Phi &  =\psi\label{C3}\\
\Psi &  =2\phi \label{C4} 
\end{align}
where $a \leq R < \infty, \, 0 \leq\Theta\leq\pi, \, 0 \leq\Phi\leq2\pi, \, 0
\leq\Psi\leq4\pi$, then the Eguchi-Hanson metric (\ref{EHstandard}) transforms
into the two-center Gibbons-Hawking form (\ref{Aseries})
\begin{equation}
ds^{2}=H(R,\theta)\big (dR^{2}+R^{2}(d\Theta^{2}+\sin^{2}\Theta d\Phi
^{2})\big ) +\frac{1}{H(R,\Theta)}(\frac{a}{8}d\Psi+Y(R,\theta) d\Phi)^{2}
\end{equation}
where
\begin{equation}
H(R,\Theta)=\frac{a}{8}\{ \frac{1}{R-R_{1}}+ \frac{1}{R-R_{2}} \} \label{HEH} 
\end{equation}
and
\begin{equation}
Y(R,\theta)=\frac{a}{8} \big ( \frac{R\cos\theta-a}{\sqrt{R^{2}+a^{2} 
-2Ra\cos\Theta}}+ \frac{R\cos\theta+2c}{\sqrt{R^{2}+a^{2}+2Ra\cos\Theta}}
\big ). \label{YEH} 
\end{equation}

In equations (\ref{HEH}) and (\ref{YEH}), $R_{1}=(0,0,a)$ and $R_{2}=-R_{1}$
are Euclidean position vectors of two nut singularities.

Here we consider the extension of metrics (\ref{Aseries}) by considering
\begin{equation}
V_{\epsilon}=\epsilon+\sum_{i=1}^{k} \frac{m_{i}}{\mid\vec x- \vec x _{i}
\mid}. \label{Vepsilon} 
\end{equation}
The hyper-Kahler metrics (\ref{Aseries}) with $V_{\epsilon}$ pose a
translational self-dual (or anti-self-dual) Killing vector $K_{\mu}$, that
means
\begin{equation}
\nabla_{\mu}K_{\nu}=\pm\frac{1}{2}\sqrt{det \, g}\epsilon_{\mu\nu} 
^{\rho\lambda}\nabla_{\rho}K_{\lambda}. \label{transkill} 
\end{equation}
This (anti-) self-duality condition (\ref{transkill}) implies the
three-dimensional Laplace equation for $V_{\epsilon}$ with solutions
(\ref{Vepsilon}). For $\epsilon\neq 0$ in (\ref{Vepsilon}), the metrics
(\ref{Aseries}) describe the asymptotically locally flat (ALF) multi Taub-NUT spaces. The removal of nut
singularities implies $m_{i}=m$ and $t$ a periodic coordinate of period
$\frac{8\pi m}{k}$.

\section{M2 Solutions Over Gibbons-Hawking Space}

{\label{sec:MbraneGH} }

\begin{figure}[tbp]
\centering           
\begin{minipage}[c]{.6\textwidth}
        \centering
        \includegraphics[width=\textwidth]{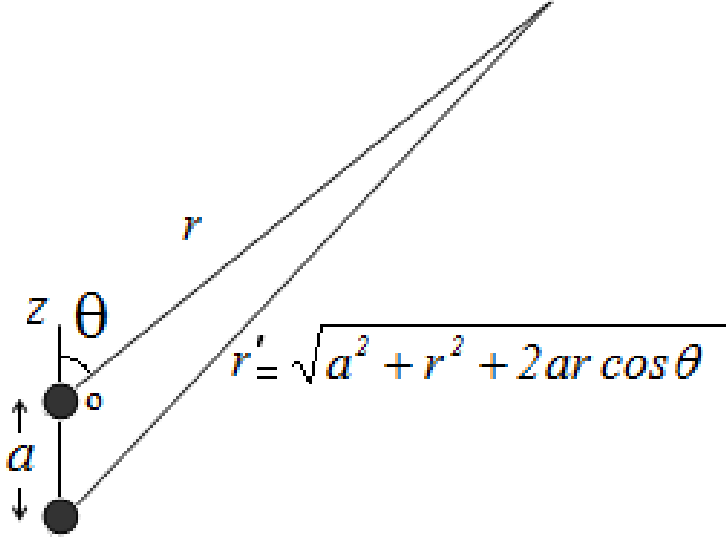}
    \end{minipage}
\caption{The geometry of charges.
}
\label{geometryofcharges}
\end{figure}

In this section, we consider the Gibbons-Hawking space with $k=2$ and metric
function $V_{\epsilon}$ with $\epsilon \neq 0$, as a part of transverse space to M2 and M5-branes. The
four-dimensional Gibbons-Hawking metric is
\begin{equation}
ds_{GH}^{2}={V_\epsilon(r,\theta)}\{dr^{2}+r^{2}(d\theta^{2}+\sin^{2}\theta d\phi^{2} 
)\}+\frac{(d\psi+\omega(r,\theta)d\phi)^{2}}{V_\epsilon(r,\theta)} 
\label{dsGH}
\end{equation}
where
\begin{align}
\omega(r,\theta)  &  =n_{1}\cos\theta+\frac{n_{2}(a+r\cos\theta)}{\sqrt
{r^{2}+a^{2}+2ar\cos\theta}}\label{om}\\
V_\epsilon(r,\theta)  &  =\epsilon+\frac{n_{1}}{r}+\frac{n_{2}}{\sqrt{r^{2} 
+a^{2}+2ar\cos\theta}}.
\label{Vep}
\end{align}
The eleven dimensional M2-brane with an embedded transverse Gibbons-Hawking
space is given by the following metric 
\begin{equation}
ds_{11}^{2}=H(y,r,\theta)^{-2/3}\left(  -dt^{2}+dx_{1}^{2}+dx_{2}^{2}\right)
+H(y,r,\theta)^{1/3}\left(  dy^{2}+y^{2}d\Omega_{3}^{2}+ds_{GH}^{2}\right)
\label{ds11m2} 
\end{equation}
and non-vanishing four-form field components are given by eqs. (\ref{Fy}), (\ref{Fr}) and
(\ref{Ft}).
The metric (\ref{ds11m2}) is a solution to the eleven dimensional supergravity
equations provided $H\left(  y,r,\theta\right)  $ is a solution to the
differential equation
\begin{align}
&  2ry\sin\theta \frac{\partial H}{\partial r}+y\cos\theta
\frac{\partial H}{\partial\theta}+r^{2}y\sin\theta \frac{\partial^{2}H}{\partial r^{2}}+y\sin\theta \frac
{\partial^{2}H}{\partial\theta^{2}}+\nonumber\\
&  +(r^{2}y\sin\theta\frac{\partial^{2}H}{\partial y^{2}}+3r^{2}\sin\theta
\frac{\partial H}{\partial\theta})V(r,\theta)=0. \label{LapH} 
\end{align}
We notice that solutions to the harmonic equation (\ref{LapH}) determine the M2-brane metric function everywhere except at the location of the brane source. To maximize the symmetry of the problem, hence simplify the analysis, we consider the M2-brane source is placed at the point $y=0,r=0$. 
Substituting
\begin{equation}
H(y,r,\theta)=1+Q_{M2}Y(y)R(r,\theta) \label{Hyrsep} 
\end{equation}
where $Q_{M2}$ is the charge on the M2-brane, we arrive at two differential
equations for $Y(y)$ and $R(r,\theta).$ The solution of the differential
equation for $Y(y)$ is
\begin{equation}
Y(y)\sim\frac{J_{1}(cy)}{y} \label{Y1} 
\end{equation}
which has a damped oscillating behavior at infinity. The differential equation
for $R(r,\theta)$ is
\begin{equation}
2r\frac{\partial R(r,\theta)}{\partial r}+r^{2}\frac{\partial
^{2}R(r,\theta)}{\partial r^{2}}+\frac{\cos\theta}{\sin\theta}\frac{\partial
R(r,\theta)}{\partial\theta}+\frac{\partial^{2}R(r,\theta)}{\partial^{2} 
\theta}=c^{2}r^{2}V(r,\theta)R(r,\theta) 
\label{a0} 
\end{equation}
where $c$ is the separation constant. First, we are interested in the
solutions of (\ref{a0}) far enough from the locations of NUT charges. So, we
take $r>>a$, hence we have $\frac{1}{r^{\prime}}=\frac{1}{r\sqrt{(\frac{a} 
{r})^{2}+1+2(\frac{a}{r})\cos\theta}}\approx\sum\limits_{l=0}P_{l}(-\cos
\theta)\frac{a^{l}}{r^{l+1}}$ where $r$ and $r^{\prime}$ are the distances to
the two NUT charges $n_{1}$ and $n_{2}$, located on $z$-axis at $(0,0,0)$ and
$(0,0,-a)$ (figure \ref{geometryofcharges}). 
We keep the first two terms in the expansion of $1/r^{\prime}$, corresponding
to $l=0,1$. The differential equation (\ref{a0}) turns to\newline 
\begin{equation}
2r\frac{\partial R(r,\theta)}{\partial r}+r^{2}\frac{\partial^{2}R(r,\theta
)}{\partial r^{2}}+\frac{\cos\theta}{\sin\theta}\frac{\partial R(r,\theta
)}{\partial\theta}+\frac{\partial^{2}R(r,\theta)}{\partial^{2}\theta} 
=c^{2}r^{2}\{\epsilon+\frac{N}{r}-\frac{\tilde{n}_{2}\cos\theta}{r^{2}}\}R(r,\theta)
\label{a1} 
\end{equation}
where $N=n_{1}+n_{2}$ and $\tilde{n}_{2}=an_{2}$.

By substituting $R(r,\theta)=f(r)g(\theta)$, we find two separated
second-order differential equations, given by \newline 
\begin{equation}
r^{2}\frac{d^{2}f(r)}{dr^{2}}+2r\frac{df(r)}{dr}-c^{2}(\epsilon r^{2} 
+Nr+M)f(r)=0 \label{a2} 
\end{equation}
\newline 
\begin{equation}
\frac{d^{2}g(\theta)}{d\theta^{2}}+\frac{\cos\theta}{\sin\theta} 
\frac{dg(\theta)}{d\theta}+c^{2}(M+\tilde{n}_{2}\cos\theta)g(\theta)=0
\label{a3} 
\end{equation}
where $M$ is the second separation constant.

\begin{figure}[tbp]
\centering           
\begin{minipage}[c]{.6\textwidth}
        \centering
        \includegraphics[width=\textwidth]{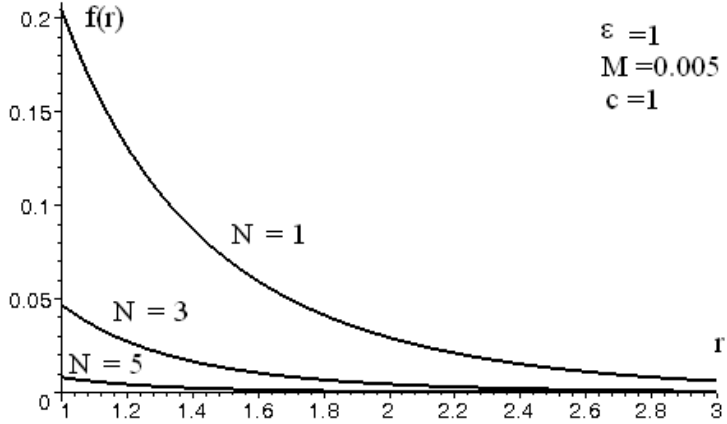}
    \end{minipage}
\caption{Solutions to eq. (\ref{a2}) with different values for N.
}
\label{figforf}
\end{figure}

\begin{figure}[tbp]
\centering           
\begin{minipage}[c]{.6\textwidth}
        \centering
        \includegraphics[width=\textwidth]{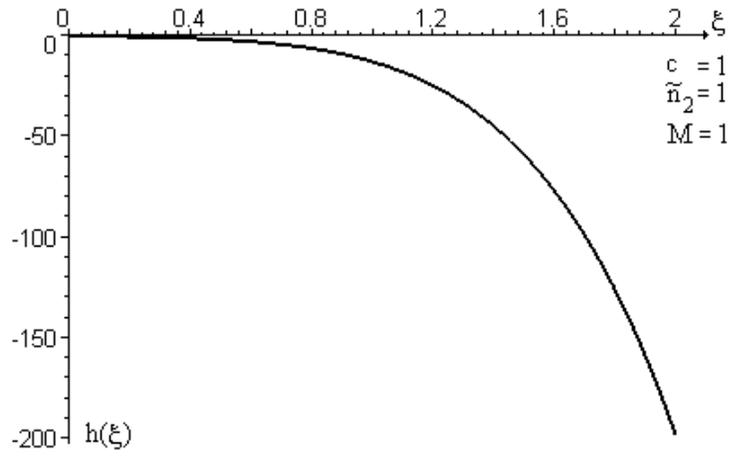}
    \end{minipage}
\caption{h($\xi$) is a well defined function around origin O.
}
\label{hofxi}
\end{figure}

We change the coordinate $r$ to $r=\frac{1}{z}$, hence the differential
equation (\ref{a2}) changes to
\begin{equation}
\frac{d^{2}f(z)}{dz^{2}}-c^{2}(\frac{\epsilon}{z^{4}}+\frac{N}{z^{3}}+\frac
{M}{z^{2}})f(z)=0. \label{a5} 
\end{equation}
The solutions to equation (\ref{a5}) are $z$ times the Whittaker functions.
So, the most general solution to (\ref{a2}) which vanishes at infinity is
\begin{equation}
f(r)=\frac{1}{r}\mathcal{W}_{W}(-\frac{cN}{2\sqrt{\epsilon}},\frac
{\sqrt{1+4Mc^{2}}}{2},2c\sqrt{\epsilon}r) \label{a6} 
\end{equation}
where $\mathcal{W}_{W}(\alpha,\beta,x)$ is the Whittaker-Watson function,
related to confluent hypergeometric function $\mathcal{U}$, by
\begin{equation}
\mathcal{W}_{W}(\alpha,\beta,x)=e^{-1/2x}x^{1/2+\beta}\mathcal{U} 
(1/2+\beta-\alpha,1+2\beta,x). \label{WW} 
\end{equation}
In figure \ref{figforf}, the behavior of $f(r)$ is given where we choose
the separation constant $c=1$, $\epsilon=1$ and $M=0.005$, respectively. 
The solutions to equation (\ref{a3}), in terms of $\xi=1-\cos\theta$, are
given by
\begin{equation}
g(\theta)=\mathcal{H}_{C}(\xi)\{C_{1}+C_{2}\int\frac{1}{\xi(\xi-2)\mathcal{H} 
_{C}^{2}(\xi)}d\xi\} \label{r7} 
\end{equation}
where $\mathcal{H}_{C}(\xi)$ stands for $\mathcal{H}_{C}(0,0,0,2\tilde{n} 
_{2}c^{2},-c^{2}(M+\tilde{n}_{2}),\frac{1}{2}\xi)$; the Heun-$C$ function. The Heun-C differential equation and functions $\mathcal{H}_{C}(\alpha,\beta,\gamma,\delta,\lambda,x)$ are reviewed briefly in
appendix A.
The first part of (\ref{r7}) which is proportional to $\mathcal{H}_{C}(\xi)$,
is an analytical function at $\xi=0$. However the second part of (\ref{r7}) is
not an analytical function at $\xi=0$.
To understand better the behavior of the second part of solution (\ref{r7}),
we consider the function
\begin{equation}
h(\xi)=\frac{1}{(\xi-2)\mathcal{H}_{C}^{2}(\xi)} \label{r9} 
\end{equation}
and use the Maclaurin's theorem, we get a power series expansion as
\begin{equation}
h(\xi)=\sum_{n=0}^{\infty}a_{n}\xi^{n}=a_{0}+a_{1}\xi+a_{2}\xi^{2}+\cdot
\cdot\cdot. \label{r11} 
\end{equation}
Here, the first few coefficients are given by
\begin{align}
a_{0}  &  =-\frac{1}{2}\\
a_{1}  &  =-\frac{M+\tilde{n}_{2}}{2}c^{2}-\frac{1}{4}\\
a_{2}  &  =-\frac{3Mc^{2}}{8}-\frac{5M^{2}c^{4}}{16}-\frac{5M\tilde{n} 
_{2}c^{4}}{8}-\frac{5\tilde{n}_{2}^{2}c^{4}}{16}-\frac{\tilde{n}_{2}c^{2}} 
{4}-\frac{1}{8}. 
\end{align}
In figure \ref{hofxi}, for instance the plot of $h(\xi)$\ versus $\xi$ is
given where we set $c=1$, $\tilde{n}_{2}=1$ and $M=1$. In this figure,
$h(\xi)$ is expanded up to order of $\xi^{5}$ as\newline 
\begin{equation}
h(\xi)=-\frac{1}{2}-\frac{5}{4}\xi-2\xi^{2}-\frac{191}{72}\xi^{3}-\frac
{7345}{2304}\xi^{4}-\frac{415937}{115200}\xi^{5}+O(\xi^{6}). \label{r10} 
\end{equation}
The series expansion (\ref{r11}) yields the final form of the solution
$g(\theta)$ as
\begin{equation}
g(\theta)=\mathcal{H}_{C}(1-\cos\theta)\{C_{c,M}-\frac{1}{2}C_{c,M}^{\prime
}\big (\ln(1-\cos\theta)-(\frac{M+\tilde{n}_{2}}{2}c^{2}+\frac{1}{4} 
)(1-\cos\theta)+\cdots\newline\big )\} \label{r14} 
\end{equation}
\newline or \newline 
\begin{eqnarray}
g(\theta)  &=&C_{c,M}\{\ (1-(\frac{M+\tilde{n}_{2}}{2}c^{2})(1-\cos
\theta)+(\frac{M^{2}c^{4}}{16}+\frac{Mc^{4}\tilde{n}_{2}}{8}+\frac{c^{4} 
\tilde{n}_{2}^{2}}{16}-\frac{Mc^{2}}{8})(1-\cos\theta)^{2}+
\nonumber\\
&+&O(1-\cos\theta
)^{3})\}+\nonumber\\
&+&C_{c,M}^{\prime}\{\ \ln(1-\cos\theta)\{  1-\frac{M+\tilde{n}_{2}} 
{2}c^{2})(1-\cos\theta)+(\frac{M^{2}c^{4}}{16}+\frac{Mc^{4}\tilde{n}_{2}} 
{8}+\frac{\tilde{n}_{2}^{2}c^{4}}{16}-\frac{Mc^{2}}{8})(1-\cos\theta
)^{2} +\nonumber \\
&+&O(1-\cos\theta)^{3}\}+ \nonumber\\
&+&\{  (\frac{1}{2}+(M+\tilde{n}_{2})c^{2})(1-\cos\theta)+(\frac{Mc^{2} 
}{8}+\frac{1}{8}-\frac{3c^{4}\tilde{n}_{2}^{2}}{16}-\frac{3\tilde{n}_{2} 
Mc^{4}}{8}-\frac{3M^{2}c^{4}}{16})(1-\cos\theta)^{2}+ \nonumber \\
&+& O(1-\cos\theta
)^{3}\}  \}.
\end{eqnarray}
\begin{figure}[tbp]
\centering           
\begin{minipage}[c]{.6\textwidth}
        \centering
        \includegraphics[width=\textwidth]{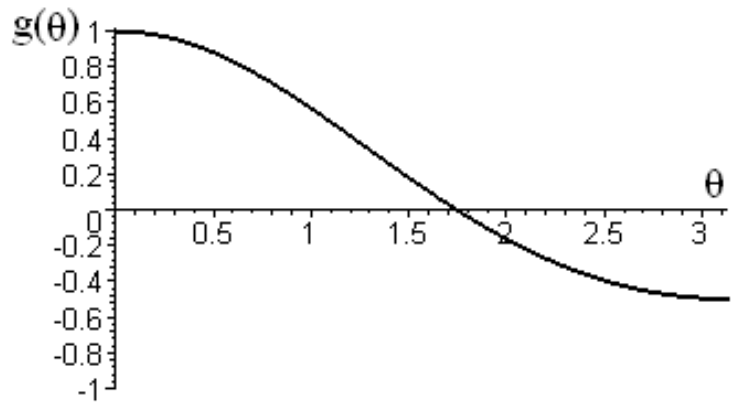}
    \end{minipage}
\caption{The graph of g($\theta$) keeping two terms of the series.
}
\label{gfun}
\end{figure}
The constant $C_{c,M}^{\prime}$ should be chosen zero, otherwise we get
logarithmic divergence at $\theta=0$ for $r>>a$. 
A typical functional form of $g(\theta)$ is shown in figure \ref{gfun}, where we set $\tilde{n}_{2}=M=c=C_{c,M}=1$ and $C'_{c,M}=0$. 
So, the solution to the
differential equation (\ref{a1}) or the asymptotic solution to (\ref{a0}) is
\begin{equation}
R(r,\theta)=\frac{C_{c,M}}{r}\mathcal{W}_{W}(-\frac{cN}{2\sqrt{\epsilon} 
},\frac{\sqrt{1+4Mc^{2}}}{2},2c\sqrt{\epsilon}r)\mathcal{H}_{C} 
(0,0,0,2\tilde{n}_{2}c^2,-(M+\tilde{n}_{2})c^{2},\frac{1}{2}\xi).\ \label{r15} 
\end{equation}
Turning next to find the exact solution to (\ref{a0}), we change the
coordinates $r,\theta$ to $\mu,\lambda$, defined by 
\begin{align}
\mu &  =r^{\prime}+r\label{r17}\\
\lambda &  =r^{\prime}-r \label{r17b} 
\end{align}
where $\mu\geqslant a$ and $-a  \leq  \lambda  \leq a$.
We notice that the coordinate transformations (\ref{r17}) and (\ref{r17b}) are
well defined everywhere except along the z-axis.
The differential equation (\ref{a0}), in the new coordinates, turns out to be
\begin{equation}
-2\lambda\frac{\partial R}{\partial\lambda}+(a^{2}-\lambda^{2})\frac
{\partial^{2}R}{\partial\lambda^{2}}+2\mu\frac{\partial R}{\partial\mu} 
+(\mu^{2}-a^{2})\frac{\partial^{2}R}{\partial\mu^{2}}=c^{2}\left[  \frac{1} 
{4}\epsilon(\mu^{2}-\lambda^{2})+\frac{1}{2}\mu(n_{1}+n_{2})+\frac{1} 
{2}\lambda(n_{1}-n_{2})\right]  R.\label{r19} 
\end{equation}
This equation is separable and yields
\begin{align}
2\lambda\frac{1}{G}\frac{\partial G}{\partial\lambda}+(\lambda^{2}-a^{2} 
)\frac{1}{G}\frac{\partial^{2}G}{\partial\lambda^{2}}-\frac{1}{2}c^{2} 
(n_{2}-n_{1})\lambda-\frac{1}{4}\epsilon c^{2}\lambda^{2}-Mc^{2} &
=0\label{r21}\\
2\mu\frac{1}{F}\frac{\partial F}{\partial\mu}+(\mu^{2}-a^{2})\frac{1}{F} 
\frac{\partial^{2}F}{\partial\mu^{2}}-\frac{1}{2}c^{2}(n_{1}+n_{2})\mu
-\frac{1}{4}\epsilon c^{2}\mu^{2}-Mc^{2} &  =0\label{r22} 
\end{align}
upon substituting in $R(\mu,\lambda)=F(\mu)G(\lambda)$ where $M$ is the
separation constant.
The solution to equation (\ref{r21}) is given by
\begin{equation}
G(\lambda)=\tilde{\mathcal{H}}_{C}(\lambda)\{\hat{g}_{c,M}+\hat{g} 
_{c,M}^{\prime}\int\frac{1}{(a-\lambda)(a+\lambda)\tilde
{\mathcal{H}}_{C}^{2}(\lambda)}d\lambda\}\label{r24} 
\end{equation}
where $\tilde{\mathcal{H}}_{C}(\lambda)$ stands for
\begin{equation}
\tilde{\mathcal{H}}_{C}(\lambda)=e^{\frac{c}{2}\sqrt{\epsilon}(a-\lambda
)}{\mathcal{H}}_{C}(2ca\sqrt{\epsilon},0,0,ac^{2}N_{-},-\frac{1}{4}(\epsilon
a^{2}+2aN_{-}+4M)c^{2},\frac{1}{2}(1-\frac{\lambda}{a})).\label{r25} 
\end{equation}
In equations (\ref{r24}) and (\ref{r25}), $N_{-}=n_{2}-n_{1}$ and $\tilde
{g}_{c,M},\tilde{g}_{c,M}^{\prime}$ are two constants in $\lambda$.
The power series expansion of $\tilde{\mathcal{H}}_{C}(\lambda)$ is \textbf{
}
\begin{align}
\tilde{\mathcal{H}}_{C}(\lambda) &  =1-(\frac{aN_{-}c^{2}}{4}+\frac{Mc^{2}} 
{2}+\frac{\epsilon a^{2}c^{2}}{8})(1-\frac{\lambda}{a})+\nonumber\\
&+ (\frac{\epsilon
a^{2}c^{2}}{32}-\frac{Mc^{2}}{8}+\frac{\epsilon^{2}a^{4}c^{4}}{256} 
+\frac{\epsilon a^{3}c^{4}N_{-}}{64}+\frac{c^{4}M^{2}}{16}+\frac{\epsilon
a^{2}c^{4}M}{32}+\frac{ac^{4}N_{-}M}{16}+\nonumber \\
&  +\frac{a^{2}c^{4}N_{-}^{2}}{64})(1-\frac{\lambda}{a})^{2}+O(\lambda^{3}).\label{r26}
\end{align}
Hence we obtain
\begin{equation}
G(\lambda)=\tilde{\mathcal{H}}_{C}(\lambda)\{g_{c,M}+g_{c,M}^{\prime} 
\ln\left\vert 1-\frac{\lambda}{a}\right\vert \}+g_{c,M}^{\prime} 
\sum_{n=1}^{\infty}d_{n}(1-\frac{\lambda}{a})^{n}
\label{r27} 
\end{equation}
\newline where ${g}_{c,M},{g}_{c,M}^{\prime}$ and $d_{n}$'s are constants in
$\lambda$. The first few $d_{n}$'s are
\begin{align}
d_{1} &  =\frac{1}{2}+Mc^{2}+\frac{\epsilon a^{2}c^{2}}{4}+\frac
{aN_{-}c^{2}}{2}\nonumber\\
d_{2} &  =\frac{Mc^{2}}{8}-\frac{\epsilon a^{2}c^{2}}{32}+\frac{1}{8} 
-\frac{3\epsilon^{2}a^{4}c^{4}}{256}-\frac{3\epsilon a^{3}c^{4}N_{-}} 
{64}- \nonumber \\
&-\frac{3c^{4}M^{2}}{16}-\frac{3\epsilon a^{2}c^{4}M}{32}-\frac
{3ac^{4}N_{-}M}{16}-\frac{3a^{2}c^{4}N_{-}^{2}}{64}. 
\label{dcoeffs}
\end{align}
The same approach can be used to find the solution to
equation (\ref{r22}). We find
\begin{equation}
F(\mu)=\tilde{\mathcal{H}}_{C}(\mu)\{\hat{f}_{c,M}+\hat{f}_{c,M}^{\prime
}(\mu)\int\frac{1}{(\mu -a)(a+\mu)\tilde{\mathcal{H}}_{C}^{2}(\mu
)}d\mu\}\label{r28} 
\end{equation}
where $\tilde{\mathcal{H}}_{C}(\mu)$ stands for
\begin{equation}
\tilde{\mathcal{H}}_{C}(\mu)=e^{\frac{c}{2}\sqrt{\epsilon}(a-\mu)} 
{\mathcal{H}}_{C}(2ca\sqrt{\epsilon},0,0,ac^{2}N_{+},-\frac{1}{4}(\epsilon
a^{2}+2aN_{+}+4M)c^{2},\frac{1}{2}(1-\frac{\mu}{a})).\label{r29} 
\end{equation}
In equation (\ref{r29}), $N_{+}=n_{1}+n_{2}$ which yields the power series
expansion as \newline 
\begin{align}
\tilde{\mathcal{H}}_{C}(\mu) &  =1-(\frac{aN_{+}c^{2}}{4}+\frac{Mc^{2}} 
{2}+\frac{\epsilon a^{2}c^{2}}{8})(1-\frac{\mu}{a})+\nonumber \\
&+(\frac{\epsilon
a^{2}c^{2}}{32}-\frac{Mc^{2}}{8}+\frac{\epsilon^{2}a^{4}c^{4}}{256} 
+\frac{\epsilon a^{3}c^{4}N_{+}}{64}+\frac{c^{4}M^{2}}{16}+\frac{\epsilon
a^{2}c^{4}M}{32}+\frac{ac^{4}N_{+}M}{16}+\nonumber \\
&  +\frac{a^{2}c^{4}N_{+}^{2}}{64})(1-\frac{\mu}{a})^{2}+O(\mu^{3}).
\label{r30}
\end{align}
So, we obtain
\newline 
\begin{equation}
F(\mu)=\tilde{\mathcal{H}}_{C}(\mu)\{f_{c,M}+f_{c,M}^{\prime}\ln
\left\vert 1-\frac{\mu}{a}\right\vert \}+f_{c,M}^{\prime}\sum_{n=1}^{\infty
}b_{n}(1-\frac{\mu}{a})^{n}
\label{r31} 
\end{equation}
where $b_n$'s are given by (\ref{dcoeffs}) upon replacing $N_-$ by $N_+$. In addition to the asymptotic solution, given by (\ref{r15}) for far-zone $r>>a$, as well as the solution near NUT charges (near-zone), given by (\ref{r27}) and (\ref{r31}), we can obtain the solution to equation (\ref{a0}) (or (\ref{r19})) in intermediate-zone for any values of $r$ and $\theta$ (or any values of $\mu$ and $\lambda$). The 
form of our intermediate-zone looks like the last summation term in (\ref{r27}) or (\ref{r31}). Hence, we find the most general solution to equation (\ref{r19}) (or
equivalently to equation (\ref{a0}) after coordinate transformations
(\ref{r17}) and (\ref{r17b}))
\begin{eqnarray}
R(r,\theta)&=&\left\{
\tilde{\mathcal{H}}_{C}(\mu)\{f_{c,M}+f_{c,M}^{\prime}\ln\left\vert
1-\frac{\mu}{a}\right\vert \}\delta_{a,\mu_0}+f_{c,M}^{\prime}\sum_{n=0}^{\infty}b_{n,\mu_0} 
(1-\frac{\mu}{\mu_0})^{n}\right\}\times \nonumber \\
&\times& \left\{  \tilde{\mathcal{H}}_{C}(\lambda)\{g_{c,M}+g_{c,M} 
^{\prime}\ln\left\vert 1-\frac{\lambda}{a}\right\vert \}\delta_{a,\lambda_0}+g_{c,M}^{\prime
}\sum_{n=0}^{\infty}d_{n,\lambda_0}(1-\frac{\lambda}{\lambda_0})^{n}\right\}
\label{a32} 
\end{eqnarray}
where
\begin{align}
\mu &  =\sqrt{r^{2}+a^{2}+2ar\cos\theta}+r\\
\lambda &  =\sqrt{r^{2}+a^{2}+2ar\cos\theta}-r\label{r33} 
\end{align}
and $\mu _{0}\geq a$,\,$\vert \lambda _{0}\vert \leq a$.
In (\ref{a32}), $d_{0,a}=0$ and $d_{n>0,a}$ are given by (\ref{dcoeffs}). The other coefficients are listed in appendix B. In figures 4.5 and 4.6, we plot the slices of the most general solution (\ref{a32}) at $\lambda=$const.  and $\mu=$const. respectively, for different values of separation constant $c$. 
\begin{figure}[tbp]
\centering           
\begin{minipage}[c]{.6\textwidth}
        \centering
        \includegraphics[width=\textwidth]{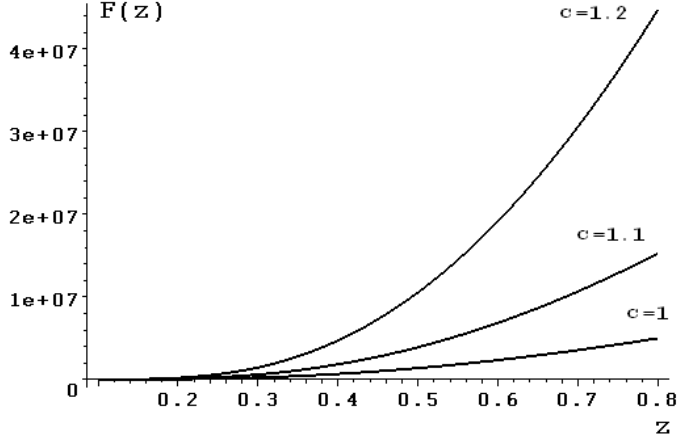}
    \end{minipage}
\caption{The first bracket in (\ref{a32}) as a function of $\mu -a=\frac{1}{z}$. 
}
\label{secsol1}
\end{figure}
\begin{figure}[tbp]
\centering           
\begin{minipage}[c]{.6\textwidth}
        \centering
        \includegraphics[width=\textwidth]{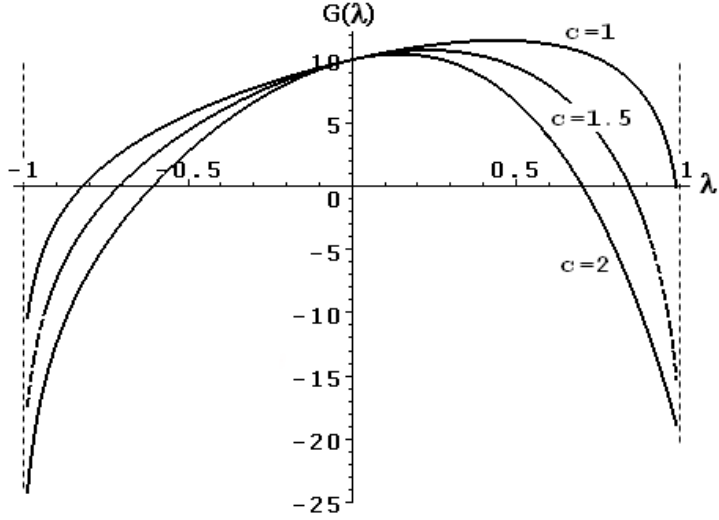}
    \end{minipage}
\caption{The second bracket in (\ref{a32}) as a function of $\lambda$. 
}
\label{secsol2}
\end{figure}

\begin{figure}[tbp]
\centering           
\begin{minipage}[c]{.6\textwidth}
        \centering
        \includegraphics[width=\textwidth]{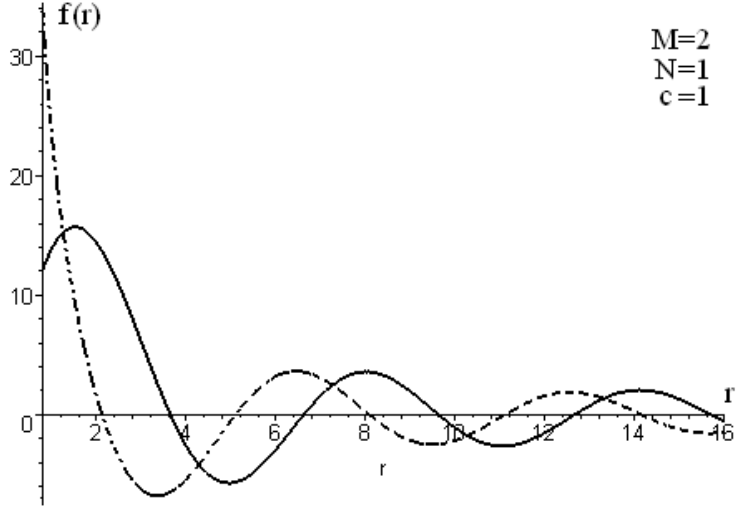}
    \end{minipage}
\caption{Two independent solutions in (\ref{v4.1}) at fixed $\xi$.
}
\label{secsol}
\end{figure}

Moreover, in addition to the general solution (\ref{a32}), we can easily obtain another
independent solution by changing the separation constant
$c$ to $ic$ in equations
(\ref{a2}) and (\ref{a3}). In this case, we have\newline 
\begin{equation}
\frac{d^{2}f(z)}{dz^{2}}+c^{2}(\frac{\epsilon}{z^{4}}+\frac{N}{z^{3}}-\frac
{M}{z^{2}})f(z)=0\, \label{rr1} 
\end{equation}
\newline 
\begin{equation}
\frac{d^{2}g(\theta)}{d\theta^{2}}+\frac{\cos\theta}{\sin\theta} 
\frac{dg(\theta)}{d\theta}+c^{2}(M-\tilde{n}_{2}\cos\theta)g(\theta)=0
\label{rr2} 
\end{equation}
\newline 
where we changed $M$ to $-M$ for convenience and $z=\frac{1}{r}$. The second solution then, is given by
\begin{eqnarray}
\tilde R(r,\theta)&=&\frac{1}{r}\left\{  C_{c,W}\mathcal{W}_W(-\frac{icN}{2\sqrt
{\epsilon}},\frac{\sqrt{1+4Mc^{2}}}{2},2ic\sqrt{\epsilon}r)+C_{c,M} 
\mathcal{W}_{M}(-\frac{icN}{2\sqrt{\epsilon}},\frac{\sqrt{1+4Mc^{2}}} 
{2},2ic\sqrt{\epsilon}r)\right\}\times\nonumber\\&\times& {\mathcal{H}}_{C}(0,0,0,-2\tilde{n}_{2}c^2,-(M-\tilde
{n}_{2})c^{2},\frac{1}{2}\xi).\ 
\label{v4.1} 
\end{eqnarray}
In figure \ref{secsol}, the solution (\ref{v4.1}) at a constant $\xi$ has been plotted. 
The most general solution to equation (\ref{a0}) after analytic continuation of $c$ is
given by $\tilde R(r,\theta)=\tilde F(\mu) \tilde G(\lambda)$. We find
\begin{equation}
\tilde G(\lambda)=\tilde{\tilde{\mathcal{H}}}_{C}(\lambda)\{\hat{\tilde{g}}_{c,M}+\hat{\tilde{g}} 
_{c,M}^{\prime}\int\frac{1}{(a-{\lambda})(a+\lambda)\tilde{\tilde
{\mathcal{H}}}_{C}(\lambda)}d\lambda\}
\end{equation}
\newline 
where $\tilde{\tilde{\mathcal{H}}}_{C}(\lambda)$ stands for
\begin{equation}
\tilde{\tilde{\mathcal{H}}}_{C}(\lambda)=e^{\frac{ic}{2}\sqrt{\epsilon}(a-\lambda
)}{\mathcal{H}}_{C}(2ica\sqrt{\epsilon},0,0,-ac^{2}N_{-},\frac{1}{4}(\epsilon
a^{2}+2aN_{-}-4M)c^{2},\frac{1}{2}(1-\frac{\lambda}{a}))
\end{equation}
and finally we obtain 
\begin{equation}
\tilde G(\lambda)=\tilde{\tilde{\mathcal{H}}}_{C}(\lambda)\{\tilde{g}_{c,M}+\tilde{g}_{c,M}^{\prime} 
\ln\left\vert 1-\frac{\lambda}{a}\right\vert \}+\tilde{g}_{c,M}^{\prime} 
\sum_{n=1}^{\infty}\tilde{d}_{n}(1-\frac{\lambda}{a})^{n}.
\label{Gtilde}
\end{equation}
In (\ref{Gtilde}), $\tilde{{g}}_{c,M},\tilde{{g}}_{c,M}^{\prime}$ and $\tilde{d}_{n}$'s are constants. The
first few $\tilde{d}_{n}$'s are
\begin{align}
\tilde{d}_{1} &  =\frac{1}{2}+Mc^{2}-\frac{\epsilon a^{2}c^{2}}{4}-\frac
{aN_{-}c^{2}}{2}\nonumber\\
\tilde{d}_{2} &  =\frac{Mc^{2}}{8}+\frac{\epsilon a^{2}c^{2}}{32}+\frac{1}{8} 
-\frac{3\epsilon^{2}a^{4}c^{4}}{256}-\frac{3\epsilon a^{3}c^{4}N_{-}} 
{64}-\frac{3c^{4}M^{2}}{16}+\frac{3\epsilon a^{2}c^{4}M}{32}+\frac
{3ac^{4}N_{-}M}{16}-\frac{3a^{2}c^{4}N_{-}^{2}}{64}. 
\end{align}

By the same method, we can find the function $\tilde F(\mu)$, hence we get the most general solution as
\begin{eqnarray}
\tilde{R}(r,\theta)&=&\left\{
\tilde{\tilde{\mathcal{H}}}_{C}(\mu)\{\tilde{f}_{c,M}+\tilde{f}_{c,M}^{\prime}\ln\left\vert
1-\frac{\mu}{a}\right\vert \}\delta_{a,\mu_0}+\tilde{f}_{c,M}^{\prime}\sum_{n=0}^{\infty}\tilde{b}_{n,\mu_0} 
(1-\frac{\mu}{\mu_0})^{n}\right\} \times\nonumber \\
&\times& \left\{  \tilde{\tilde{\mathcal{H}}}_{C}(\lambda)\{\tilde{g}_{c,M}+\tilde{g}_{c,M} 
^{\prime}\ln\left\vert 1-\frac{\lambda}{a}\right\vert \}\delta_{a,\lambda_0}+\tilde{g}_{c,M}^{\prime
}\sum_{n=0}^{\infty}\tilde{d}_{n,\lambda_0}(1-\frac{\lambda}{\lambda _0})^{n}\right\}.
\label{a322} 
\end{eqnarray}
We should note that the $y$ dependence of M2-brane metric function is 
\begin{equation}
\tilde Y(y) \sim \frac{K_1(cy)}{y}.
\end{equation}
So, the second M2-brane metric function is 
\begin{equation}
\tilde H(y,r,\theta)=1+Q_{M2}\int _0^\infty dc \int_0^\infty\,dM\tilde Y (y)\tilde R(r,\theta).
\label{secsolM2}
\end{equation}

We consider now the Gibbons-Hawking space with $k=2$ (\ref{dsGH}) with $\epsilon=0$ in (\ref{Vep}) (or equivalently the metric (\ref{EHstandard})). We should mention that despite some numerical solutions for the M-brane metric function (with embedded Eguchi-Hanson transverse metric (\ref{EHstandard})) have been found in \cite{CGMM2}, the exact closed analytic form for the M-brane function hasn't yet been found. Our method in this paper allows to construct the exact solutions for the M-brane function with embedded Eguchi-Hanson space. In the limit of $r>>a$, the solution to (\ref{a2}) (with $\epsilon=0$) is given by
\begin{equation}
f(r)=\frac{f_{c,M}K_{\sqrt{1+4Mc^{2}}}\left(2c\sqrt{Nr} 
\right)}{\sqrt r}\label{EpsilonZ1} 
\end{equation}
where $N=n_{1}+n_{2}$, in exact agreement with the numerical result of \cite{CGMM2}. The exact M-brane function is given by equation (\ref{a32}) where $\epsilon=0$ should be considered in $\tilde{\mathcal{H}}_{C}(\lambda), \tilde{\mathcal{H}}_{C}(\mu),f_{c,M},$ $g_{c,M},f'_{c,M},g'_{c,M}, d_n$ and $b_m$. Changing $c$ to $ic$ generates the second set of solutions that in the limit of $r>>a$ yields
\begin{equation}
\tilde {f}(r)=\frac{\tilde{f}_{c,M}J_{\sqrt{1+4Mc^{2}}}(2c\sqrt
{Nr})+\tilde{f}_{c,M}^{\prime}Y_{\sqrt{1+4Mc^{2}}}(2c\sqrt{Nr})}{\sqrt r}.
\label{EpsilonZ2} 
\end{equation}
We note that the general solution of the metric function could be written as a superposition of the solutions with separation constants $c$ and $M$. For example, the general first set of solution (corresponding to embedded Gibbons-Hawking space with $k=2$ and $\epsilon\neq 0$) is
\begin{eqnarray}
H(y,r,\theta)&=&1+Q_{M2}\int_0^\infty dc \int_0^\infty dM\, \frac{J_1(cy)}{y}\times\nonumber\\
&\times&
\left\{
\tilde{\mathcal{H}}_{C}(\mu)\{f_{c,M}+f_{c,M}^{\prime}\ln\left\vert
1-\frac{\mu}{a}\right\vert \}\delta_{a,\mu_0}+f_{c,M}^{\prime}\sum_{n=0}^{\infty}b_{n,\mu_0} 
(1-\frac{\mu}{\mu_0})^{n}\right\}
\times \nonumber \\
&\times& 
\left\{  \tilde{\mathcal{H}}_{C}(\lambda)\{g_{c,M}+g_{c,M} 
^{\prime}\ln\left\vert 1-\frac{\lambda}{a}\right\vert \}\delta_{a,\lambda_0}+g_{c,M}^{\prime
}\sum_{n=0}^{\infty}d_{n,\lambda_0}(1-\frac{\lambda}{\lambda_0})^{n}\right\}
.
\label{2nutmetric}
\end{eqnarray}

As we notice, the solution (\ref{2nutmetric}) depends on four combinations of constants $f_{c,M},f'_{c,M}$ and $g_{c,M},g'_{c,M}$ in form of $fg,f'g,fg'$ and $f'g'$ which each combination has dimension of inverse charge (or inverse length to six). Hence, the functional form of each constant could be considered as an expansion of the form $c^{3+2\beta}M^\beta$ where $\beta \in \mathbb{Z}_+$. Moreover we should mention the meaning of $\mu_0$ and $\lambda_0$ in equation (\ref{2nutmetric}) that have dimensions of length. We recall that the near-zone solutions (\ref{r27}) and (\ref{r31}) are given partly by series expansions around $r\simeq a$. The intermediate-zone solutions are given by similar power series expansions (with substitutions $a \rightarrow \lambda_0$ and $d_n\rightarrow d_{n,\lambda_0}$ in (\ref{r27}) and $a \rightarrow \mu_0$ and $b_n\rightarrow b_{n,\mu_0}$ in (\ref{r31}) around some fixed points, denoted by $\mu_0$ and $\lambda_0$. To calculate numerically the membrane metric function (\ref{2nutmetric}) at any $\mu, \lambda$ (or equivalently any $r$ and $\theta$), we consider some fixed values for $\mu_0$ and $\lambda_0$ (see appendix B). 

Dimensional reduction of M2-brane metric (\ref{ds11m2}) with the metric functions (for example 
(\ref{2nutmetric})) along the coordinate $\psi $\ of the metric (\ref{dsGH}) gives
type IIA supergravity metric 
\begin{eqnarray}
ds_{10}^{2} &=&H^{-1/2}(y,r,\theta)V_\epsilon^{-1/2}(r,\theta)\left(
-dt^{2}+dx_{1}^{2}+dx_{2}^{2}\right) +  \nonumber \\
&+&H^{1/2}(y,r,\theta)V_\epsilon^{-1/2}(r,\theta)\left( dy^{2}+y^{2}d\Omega _{3}^{2}\right)+\nonumber\\
&+&H^{1/2}(y,r,\theta)V_\epsilon^{1/2}(r,\theta)(dr^{2}+r^{2}d\Omega _{2}^{2})
\label{ds10TN4}
\end{eqnarray} 
which describes a localized D2-brane at $y=r=0$ along the world-volume of D6-brane, for any choice of constants in the form of $c^{3+2\beta}M^\beta$ where $\beta \in \mathbb{Z}_+$. The
other fields in ten dimensions are NSNS fields 
\begin{eqnarray}
\Phi  &=&\frac{3}{4}\ln \left\{ \frac{H^{1/3}(y,r,\theta)}{V_\epsilon(r,\theta)}\right\}  \\
B_{\mu \nu } &=&0
\end{eqnarray} 
and Ramond-Ramond (RR) fields 
\begin{eqnarray}
C_{\phi } &=&\omega(r,\theta) \\
A_{tx_{1}x_{2}} &=&\frac{1}{H(y,r,\theta)}.
\end{eqnarray}
The intersecting configuration is BPS since it has been obtained 
by compactification along a transverse direction from the BPS membrane solution with harmonic metric function (for example (\ref{2nutmetric})) \cite{forbps}. Moreover, in section \ref{sec:susy}, we use the Killing spinor equation (\ref{killingspinoreq}) to calculate how much supersymmetry is preserved by M2-brane solutions in eleven
dimensions. We conclude that half of the
supersymmetry is removed by the projection operator that is due to the presence of the brane,
and another half is removed due to the self-dual nature of the Gibbons-Hawking metric. Hence 
embedding any Gibbons-Hawking metric into an
eleven dimensional M2-brane metric preserves 1/4 of the supersymmetry. 

\section{M5 Solutions Over Gibbons-Hawking Space}

\label{sec:M5}

The eleven dimensional M5-brane metric with an embedded Gibbons-Hawking
metric has the following form 
\begin{eqnarray}
ds_{11}^{2}&=&H(y,r,\theta)^{-1/3}\left(
-dt^{2}+dx_{1}^{2}+dx_{2}^{2}+dx_{3}^{2}+dx_{4}^{2}+dx_{5}^{2}\right) +\nonumber \\ 
&+&H(y,r,\theta)^{2/3}\left( dy^{2}+ds_{GH}^{2}\right) 
\label{ds11m5p}
\end{eqnarray} 
with field strength components 
\begin{eqnarray}
F_{\psi \phi r y}&=&\frac{\alpha }{2}\sin (\theta )\frac{\partial H}{ 
\partial \theta} \nonumber\\ 
F_{\psi \phi \theta y}&=&-\frac{\alpha }{2}r^2\sin (\theta )\frac{ 
\partial H}{\partial r} \nonumber \\
F_{\psi \phi \theta r}&=&\frac{\alpha }{2}r^2\sin (\theta )V(r,\theta)\frac{ 
\partial H}{\partial y}.
\label{FScompo}
\end{eqnarray} 
We consider the M5-brane which corresponds to $\alpha =+1$;\ the $\alpha =-1$ case corresponds to an anti-M5 brane.

The metric (\ref{ds11m5p}) is a solution to the equations (\ref{GminGG}) and (\ref{dF}), provided $H\left( y,r,\theta\right) $ is a solution to the
differential equation 
\begin{align}
&  &2r\frac{\sin\theta}{V_\epsilon(r,\theta)}\frac{\partial H}{\partial r}+\frac{\cos\theta}{V_\epsilon(r,\theta)}\frac{\partial H}{\partial\theta}+r^2\sin\theta\frac{\partial^{2}H}{\partial y^{2}}  +\frac{\sin\theta}{V_\epsilon(r,\theta)}\{\frac{\partial^{2}H}{\partial\theta^{2}}+r^2\frac{\partial^{2}H}{\partial r^{2}}\}=0.\nonumber\\
&
\label{HeqforM5}
\end{align}
This equation is straightforwardly separable upon substituting 
\begin{equation}
H(y,r,\theta)=1+Q_{M5}Y(y)R(r,\theta)
\end{equation} 
where $Q_{M5}$ is the charge on the M5-brane. The solution to the
differential equation for $Y(y)$ is 
\begin{equation}
Y(y)=\cos (cy+\varsigma )
\end{equation} 
and the differential equation for $R(r,\theta)$ is the same equation as 
(\ref{a0}). Hence the most general M5-brane function (corresponding to embedded Gibbons-Hawking space with $k=2$ and $\epsilon\neq 0$) is given by
\begin{eqnarray}
H(y,r,\theta)=1&+&Q_{M5}\int_0^\infty dc \int_0^\infty d{M} \cos(cy+\varsigma)\times\nonumber\\
&\times&\left\{
\tilde{\mathcal{H}}_{C}(\mu)\{f_{c,M}+f_{c,M}^{\prime}\ln\left\vert
1-\frac{\mu}{a}\right\vert \}\delta_{a,\mu_0}+f_{c,M}^{\prime}\sum_{n=0}^{\infty}b_{n,\mu_0} 
(1-\frac{\mu}{\mu_0})^{n}\right\}
\times \nonumber \\
&\times& 
\left\{  \tilde{\mathcal{H}}_{C}(\lambda)\{g_{c,M}+g_{c,M} 
^{\prime}\ln\left\vert 1-\frac{\lambda}{a}\right\vert \}\delta_{a,\lambda_0}+g_{c,M}^{\prime
}\sum_{n=0}^{\infty}d_{n,\lambda_0}(1-\frac{\lambda}{\lambda_0})^{n}\right\}.
\label{M5sol1}
\end{eqnarray} 
Similar result holds for embedded Gibbons-Hawking space with $k=2$ and $\epsilon=0$. 
The solution (\ref{M5sol11}) depends on four 
combinations of constants in form of $fg,f'g,fg'$ and $f'g'$ which each combination should have dimension 
of inverse length. Hence, the functional form of each constant could be considered as 
an expansion of the form $c^{1/2+2\beta}M^\beta$ where $\beta \in \mathbb{Z}_+$.
As with M2-brane case, reducing (\ref{ds11m5p}) to ten dimensions gives the
following NSNS dilaton 
\begin{equation}
\Phi =\frac{3}{4}\ln \left\{ \frac{H^{2/3}(y,r,\theta)}{V_\epsilon(r,\theta)}\right\} .
\label{dilTN}
\end{equation} 
The NSNS field strength of the two-form associated with the NS5-brane, is given by 
\begin{equation}
\mathcal{H}_{(3)}={F_{\phi y r\psi }}d\phi
\wedge dy \wedge dr+{F_{\phi y \theta \psi}}d\phi \wedge dy \wedge d\theta
+{F_{\phi r \theta \psi }}d\phi \wedge dr \wedge d\theta
\label{tnH3}
\end{equation} 
where the different components of 4-form $F$, are given by ( 
\ref{FScompo}). The RR fields are 
\begin{eqnarray}
C_{(1)} &=&\omega(r,\theta )\label{tnRR} \\
\mathcal{A}_{\alpha \beta \gamma } &=&0  \label{tnA3}
\end{eqnarray} 
where $C_{\alpha }$ is the field associated with the D6-brane, and the
metric in ten dimensions is given by: 
\begin{eqnarray}
ds_{10}^{2} &=&V_\epsilon^{-1/2}(r,\theta)\left(
-dt^{2}+dx_{1}^{2}+dx_{2}^{2}+dx_{3}^{2}+dx_{4}^{2}+dx_{5}^{2}\right) +H(y,r,\theta) 
V_\epsilon^{-1/2}(r,\theta)dy^{2}+  \nonumber \\
&+&H(y,r,\theta)V_\epsilon^{1/2}(r,\theta)\left( dr^{2}+r^{2}d\Omega _{2}^{2}\right).
\label{tng10}
\end{eqnarray} 
From (\ref{tnH3}), (\ref{tnRR}), (\ref{tnA3}) and the metric (\ref{tng10}), 
we can see the above
ten dimensional metric is an NS5$\perp $D6(5) brane solution. We
have explicitly checked the BPS 10-dimensional metric (\ref{tng10}),
with the other fields (the dilaton (\ref{dilTN}), the 1-form field (\ref 
{tnRR}), and the NSNS field strength (\ref{tnH3})) make a solution to the
10-dimensional supergravity equations of motion. 
As we discuss in section \ref{sec:susy}, the solution (\ref{ds11m5p}) preserves 1/4 of
the supersymmetry. 

\section{M2-Branes With Two Transverse Gibbons-Hawking Spaces}

\label{sec:M2mix}

We can also embed two four dimensional Gibbons-Hawking spaces into the eleven dimensional
membrane metric. Here we consider the embedding of two double-NUT (or two double-center Eguchi-Hanson)
metrics of the form (\ref{dsGH}) with $\epsilon\neq 0$ (or $\epsilon=0$). The M-brane metric is 
\begin{equation}
ds_{11}^{2} =H(y,\alpha,r,\theta)^{-2/3}\left(
-dt^{2}+dx_{1}^{2}+dx_{2}^{2}\right) +H(y,\alpha,r,\theta)^{1/3}\left(
ds_{GH(1)}^{2}+ds_{GH(2)}^{2}\right) 
\label{tn4xtn4mtrc}
\end{equation}
where $ds_{GH(i)},\, i=1,2$ are two copies of the metric (\ref{dsGH}) with coordinates $(r,\theta,\phi,\psi)$ and $(y,\alpha,\beta,\gamma)$.
The non-vanishing components of four-form field are
\begin{equation}
F_{tx_{1}x_{2}x}=-\frac{1}{2H^{2}}\frac{\partial H(y,\alpha,r,\theta)}{\partial x}
\label{tn4xtn4F012i}
\end{equation} 
where $x=r,\theta,y,\alpha$.
The metric (\ref{tn4xtn4mtrc}) and four-form field (\ref{tn4xtn4F012i}) satisfy the eleven
dimensional equations of motion if
\begin{eqnarray}
& &2ry\sin(\alpha)\sin(\theta)\{V_\epsilon(r,\theta)y\frac{\partial H}{\partial r}+V_\epsilon(y,\alpha)r\frac{\partial H}{\partial y}\}+\nonumber\\
&+& \sin(\alpha)y^2\cos(\theta)V_\epsilon(r,\theta)\frac{\partial H}{\partial \theta}+r^2\sin(\theta)\cos(\alpha)V_\epsilon(y,\alpha)\frac{\partial H}{\partial \alpha}+ \nonumber \\
&+& r^2\sin(\alpha)y^2\sin(\theta)\{V_\epsilon(r,\theta)\frac{\partial^2 H}{\partial r^2}+V_\epsilon(y,\alpha)\frac{\partial^2 H}{\partial y^2}\}+\nonumber\\
&+&\sin(\theta)\sin(\alpha)\{r^2V_\epsilon(y,\alpha)\frac{\partial^2 H}{\partial \alpha^2}+y^2V_\epsilon(r,\theta)\frac{\partial^2 H}{\partial \theta^2}\}=0
\label{deqtnxtn}
\end{eqnarray} 
where $V_\epsilon(y,\alpha)=\epsilon+\frac{n_3}{y}+\frac{n_4}{\sqrt{y^2+b^2+2by\cos(\alpha)}}$.
The equation (\ref{deqtnxtn}) is separable if we set $ 
H(y,\alpha,r,\theta)=1+Q_{M2}R_{1}(y,\alpha)R_{2}(r,\theta)$. This gives two equations
\begin{equation}
2x_i\frac{\partial R_i}{\partial x_i}+x_i^{2}\frac{\partial
^{2}R_i}{\partial x_i^{2}}+\frac{\cos y_i}{\sin y_i}\frac{\partial
R_i}{\partial y_i}+\frac{\partial^{2}R_i}{\partial^{2} 
y_i}=u_ic^{2}x_i^{2}V_\epsilon(x_i,y_i)R_i 
\label{tn4xtn4deqRi}
\end{equation}
where $(x_1,y_1)=(y,\alpha)$ and $(x_2,y_2)=(r,\theta)$. There is no summation on index $i$ and $u_1=+1,\,u_2=-1$, in equation (\ref{tn4xtn4deqRi}).
We already know the solutions to the two differential equations (\ref{tn4xtn4deqRi}) as given by (\ref{a32}) and (\ref{a322}), hence the most general solution to (\ref{deqtnxtn}) is 
\begin{equation}
H(y,\alpha,r,\theta)=1+Q_{M2}\int_{0}^{\infty}dc \int_0^\infty dM \int_0^\infty d\tilde{M} R(y,\alpha)\tilde{R}(r,\theta).
\label{HTN4TN4}
\end{equation} 
We note that changing $c$ to $ic$\ in (\ref{tn4xtn4deqRi}) makes a second solution given by replacements $
R(y,\alpha)$ to $\tilde{R}(y,\alpha)$ and $\tilde{R}(r,\theta)$ to $R(r,\theta)$ in (\ref{HTN4TN4}). However the second solution is not independent of the first one.

We can choose to compactify down to ten dimensions by compactifying on
either $\psi$ or $\gamma$ coordinates. 
In the first case, we find the type IIA string
theory with the NSNS fields 
\begin{eqnarray}
\Phi & = & \frac{3}{4}\ln \left( \frac{H^{1/3}}{V_\epsilon(r,\theta)}\right) \label{oh1}\\ 
B_{\mu \nu } & = & 0 
\label{tnXtnNSNS}
\end{eqnarray} 
and RR fields
\begin{eqnarray}
C_{\phi} & = & \omega(r,\theta) \label{oh2}\\ 
A_{tx_{1}x_{2}} & = & H(y,\alpha,r,\theta)^{-1}. 
\label{tnXtnRR}
\end{eqnarray} 
The metric is given by 
\begin{eqnarray}
ds_{10}^{2} &=&H(y,\alpha,r,\theta)^{-1/2}{V_\epsilon(r,\theta)}^{-1/2}\left(
-dt^{2}+dx_{1}^{2}+dx_{2}^{2}\right) +
\nonumber \\
&+&H(y,\alpha,r,\theta)^{1/2}{V_\epsilon(r,\theta)}^{-1/2}\left( ds_{GH(1)}^{2}\right) +  
\nonumber \\
&+&H(y,\alpha,r,\theta)^{1/2}{V_\epsilon(r,\theta)}^{1/2}\left(
dr^{2}+r^{2}\left( d\theta^{2}+\sin ^{2}(\theta)d\phi
^{2}\right) \right).  
\label{ds10tnXtn}
\end{eqnarray} 
In the latter case, the type IIA fields are in the same form as (\ref{oh1}), (\ref{tnXtnNSNS}), (\ref{oh2}), (\ref{tnXtnRR}) and (\ref{ds10tnXtn}), just by replacements $(r,\theta,\phi,\psi) \Leftrightarrow (y,\alpha,\beta,\gamma)$.
In either cases, we get a fully localized D2/D6 brane system.
We can further reduce the metric (\ref{ds10tnXtn}) along the $\gamma$
direction of the first Gibbons-Hawking space. However the result of this
compactification is not the same as the reduction of the M-theory solution 
(\ref{tn4xtn4mtrc}) over a torus, which is compactified type IIB theory. The
reason is that to get the compactified type IIB theory, we should compactify
the T-dual of the IIA metric (\ref{ds10tnXtn}) over a circle, and not
directly compactify the 10D IIA metric (\ref{ds10tnXtn}) along the $\gamma$ direction.
We note also an interesting result in reducing the 11D metric 
(\ref {tn4xtn4mtrc}) along the $\psi$ (or $\gamma$) direction of the $ 
GH(1)$ (or $GH(2)$) in large radial coordinates. As $y$ 
(or $r$) $\rightarrow \infty $\ the transverse geometry in 
(\ref{tn4xtn4mtrc}) locally approaches $\mathbb{R}^{3}\otimes S^{1}\otimes GH(2)$
(or $GH(1)\otimes \mathbb{R}^{3}\otimes S^{1}$). Hence the reduced theory,
obtained by compactification over the circle of the Gibbons-Hawking, is IIA. Then
by T-dualization of this theory (on the remaining $S^{1}$ of the transverse
geometry), we find a type IIB theory which describes the D5 defects.
The solutions (\ref{tn4xtn4mtrc}) (with $\epsilon=0$ or $\epsilon\neq 0$) are BPS and also preserve 1/4 of the supersymmetry, as we show in the next section. 

\section{Supersymmetries of the Solutions}

\label{sec:susy}

In this section, we explicitly show all our BPS solutions presented in the previous sections preserve 1/4 of the supersymmetry.
Generically a configuration of $n$ intersecting branes preserves $\frac{1}{2^n}$ of the supersymmetry. In general, the Killing spinors are projected out by product of Gamma matrices with indices tangent to each brane. If all the projections are independent, then $\frac{1}{2^n}$-rule can give the right number of preserved supersymmetries. On the other hand, if the projections are not independent then $\frac{1}{2^n}$-rule can't be trusted. There are some important brane configurations when the number of preserved supersymmetries is more than that by  $\frac{1}{2^n}$-rule \cite{ex,ex2}.

As we briefly mentioned in the introduction, 
the number of non-trivial solutions to the Killing spinor equation
\begin{equation}
\partial_{M}\varepsilon+\frac{1}{4}\omega_{abM}\Gamma^{ab}\varepsilon+\frac{1} 
{144}\Gamma_{M}^{\phantom{m}npqr}F_{npqr}\varepsilon-\frac{1}{18}\Gamma
^{pqr}F_{mpqr}\varepsilon=0 \label{killingspinoreq2} 
\end{equation}
determine the amount of supersymmetry of the solution where the indices $M,N,P,...$ are eleven dimensional world indices and $a,b,...$ are eleven dimensional non-coordinate tangent space indices. The connection one-form is given by $\omega ^a_b=\Gamma^a_{bc}\hat\theta^b$, in terms of Ricci rotation coefficients $\Gamma _{abc}$ and non-coordinate basis $\hat \theta^a=e^a_Mdx^M$ where $e^M_a$ are vielbeins. The eleven dimensional M-brane metrics (\ref{ds11genM2}) and (\ref{ds11general}) are $ds^2=\eta_{ab}\hat \theta ^a \otimes \hat \theta ^b$ in non-coordinate basis. The connection one-form $\omega ^a_b$ satisfies torsion- and curvature-free Cartan's structure equations
\begin{eqnarray}
d\hat \theta ^a+\omega ^a_b \wedge \hat \theta ^b &=& 0\\
d\omega^a_b+\omega ^a_c \wedge \omega ^c_b &=& 0
\end{eqnarray}
In (\ref{killingspinoreq2}), $\Gamma^{a}$ matrices make the Clifford algebra
\begin{equation}
\left\{  \Gamma^{a},\Gamma^{b}\right\}  =-2\eta^{ab}. \label{cliffalg2} 
\end{equation}
and $\Gamma ^{ab}=\Gamma ^{\lbrack a}\Gamma ^{b]}$. Moreover, 
$\Gamma^{M_{1}\ldots M_{k}}=\Gamma^{\lbrack M_{1}}\ldots\Gamma^{M_{n}]}$.  
A representation of the algebra is given in appendix C. 

For our purposes, we use the thirty two dimensional representation of the Clifford algebra 
(\ref{cliffalg2}), given by \cite{gauntlettetal3} 
\begin{eqnarray}
\Gamma _{i}&=&\left[ 
\begin{array}{cc}
0 & -\widetilde{\Gamma }_{i} \\ 
\widetilde{\Gamma }_{i} & 0%
\end{array}%
\right] ~~~(i=1\ldots 8) \label{gamis}\\ 
\Gamma _{9}&=&\left[ 
\begin{array}{cc}
1 & 0 \\ 
0 & -1%
\end{array}%
\right] \\ 
\Gamma _{\star}&=&\left[ 
\begin{array}{cc}
0 & 1 \\ 
1 & 0%
\end{array}%
\right] \\ 
\Gamma _{0}&=&-\Gamma _{123456789\star}%
\label{Gammas}
\end{eqnarray}
We note $\Gamma _{0123456789\star}=\epsilon_{{0123456789\star}}=1$. For a given Majorana spinor $\epsilon$, its conjugate is given by $\bar \epsilon=\epsilon ^T\Gamma _0$.  Moreover we notice that $\Gamma_0\Gamma_{a_1a_2\cdots a_n}$ is symmetric for $n=1,2,5$ and antisymmetric for $n=0,3,4$. The $\widetilde{\Gamma }_{i}$'s in (\ref{gamis}), the sixteen dimensional
representation of the Clifford algebra in eight dimensions, are given by \cite{FF}%
\begin{eqnarray}
\widetilde{\Gamma }_{i} &=&\left[ 
\begin{array}{cc}
0 & L_{i} \\ 
L_{i} & 0%
\end{array}%
\right] ~~~(i=1\ldots 7)  \label{Gammatildas} \\
\widetilde{\Gamma }_{8} &=&\left[ 
\begin{array}{cc}
0 & -1 \\ 
1 & 0%
\end{array}%
\right]
\end{eqnarray}%
in terms of $L_{i}$, the left multiplication by the imaginary
octonions on the octonions. The imaginary unit octonions satisfy the
following relationship 
\begin{equation}
o_{i}\cdot o_{j}=-\delta _{ij}+c_{ijk}o_{k}  \label{unitoct}
\end{equation}%
where $c_{ijk}$\ is totally skew symmetric and its non-vanishing components
are given by
\begin{equation}
c_{124}=c_{137}=c_{156}=c_{235}=c_{267}=c_{346}=c_{457}=1.
\label{nonvanishingcompo}
\end{equation}%

We take the $L_{i}$ to be the matrices such that the relation (\ref{unitoct}) holds. In other words, given a vector $v=\left( v_{0},v_{i}\right) $ in 
$\mathbb{R}^{8}$, we write $\hat{v}=v_{0}+v_{j}o_{j}$, where the effect of
left multiplication is $o_{i}\left( \hat{v}\right)
=v_{0}o_{i}-v_{i}+c_{ijk}v_{j}o_{k}$ , we then construct the $8\times 8$
matrix $\left( L_{i}\right) _{\xi\zeta}$ by requiring $o_{i}\left( \hat{v}\right)
=\left( L_{i}\right) _{\xi\zeta}o_{\xi}v_{\zeta}$, where $\xi , \zeta =0,1,\ldots 7$. 
We consider first the M2-brane solutions considered in section 4, for example (\ref{2nutmetric}). Substituting $\varepsilon=H^{-1/6}\epsilon$ in the Killing spinor equations 
(\ref {killingspinoreq2}) yields solutions that\footnote{In what follows in this section, we show the non-coordinate tangent space indices of $\Gamma$'s by $t,x_1,x_2,\cdots,\phi,\psi$, to simplify the notation.}
\begin{equation}
\Gamma ^{{t}{x}_{1}{x}_{2}}\epsilon =-\epsilon
\label{proj012}
\end{equation}%
and so at most half the supersymmetry is preserved due to the presence of the brane.
We note that if we multiply all the components of four-form field strength, given in (\ref{Fy}),(\ref{Fr}) and (\ref{Ft}), by $-1$, then 
the projection equation (\ref{proj012}) changes to $\Gamma ^{{t}{x}_{1}{x}_{2}}\epsilon =+\epsilon$.
The other remaining equations in (\ref%
{killingspinoreq2}), arising from the left-over terms from $\partial _{M}\epsilon +
\frac{1}{4}\omega _{Mab}\Gamma ^{ab}\epsilon $\ portion, are 
\begin{eqnarray}
\partial _{\alpha _{1}}\epsilon &-&\frac{1}{2}\Gamma ^{{y}{\alpha}%
_{1}}\epsilon  =0  \label{tn4da1} \\
\partial _{\alpha _{2}}\epsilon &-&\frac{1}{2}\left[{\sin (\alpha _{1})}\Gamma ^{{y}%
{\alpha}_{2}}+{\cos (\alpha _{1})}\Gamma ^{{\alpha}%
_{1}{\alpha}_{2}}\right]\epsilon  =0  \label{tn4da2} \\
\partial _{\alpha _{3}}\epsilon &-&\frac{1}{2}\left[\sin (\alpha _{2})({\sin (\alpha _{1})}\Gamma ^{{y}{\alpha}_{3}}+ \cos
(\alpha _{1})\Gamma ^{{\alpha}_{1}{\alpha}_{3}})+%
\cos (\alpha _{2})\Gamma ^{{\alpha}_{2}{\alpha}_{3}}\right]\epsilon 
=0  \label{tn4da3} \\
\partial _{\psi }\epsilon &+&\frac{1}{4r^2\sin\theta}\left[-V^2(\frac{\partial \omega}{\partial \theta}
\Gamma ^{{\theta}{\phi}}+r\frac{\partial \omega}{\partial r}\Gamma ^{{r}{\phi}})+r\sin\theta(\frac{\partial V}{\partial \theta}
\Gamma ^{{\psi}{\phi}}+r\frac{\partial V}{\partial r}\Gamma ^{{\psi}{r}})\right] \epsilon  =0
\label{tn4dpsi} \\
\partial _{\theta }\epsilon &+&\frac{1}{4r\sin\theta}\left[-V\frac{\partial \omega}{\partial \theta}\Gamma ^{{\psi}{\phi}}+\frac{r\sin\theta}{V}(r\frac{\partial V}{\partial r}-2V)\Gamma ^{{r}{\theta}}\right]\epsilon  =0
\label{tn4dtheta} \\
\partial _{\phi }\epsilon &+&\frac{1}{4}\left[\frac{\partial (V\omega)}{\partial r}
\Gamma ^{{\psi}{r}}-\frac{1}{rV\sin\theta}(V^3\omega\frac{\partial \omega}{\partial r}-r^2\sin ^2\theta\frac{\partial V}{\partial r}+2rV\sin^2\theta)\Gamma ^{{r}{\phi}}\right.
\nonumber\\
&-&\left.\frac{1}{r^2V\sin\theta}(V^3\omega\frac{\partial \omega}{\partial \theta}-r^2\sin ^2\theta\frac{\partial V}{\partial \theta}+2r^2V\sin\theta\cos\theta)\Gamma ^{{\theta}{\phi}}+\frac{1}{4r}\frac{\partial (V\omega)}{\partial \theta}
\Gamma ^{{\psi}{\theta}}\right] \epsilon
=0.~~  \notag
\\
&&  \label{tn4dphi}
\end{eqnarray}%
We can solve the first three equations, (\ref{tn4da1}), (\ref{tn4da2}) and (\ref{tn4da3}) by using the Lorentz
transformation 
\begin{equation}
\epsilon =\exp \left\{ \frac{\alpha _{1}}{2}\Gamma ^{{y}{\alpha}%
_{1}}\right\} \exp \left\{ \frac{\alpha _{2}}{2}\Gamma ^{{\alpha}_{1}%
{\alpha}_{2}}\right\} \exp \left\{ \frac{\alpha _{3}}{2}\Gamma ^{{%
\alpha}_{2}{\alpha}_{3}}\right\} \eta.  \label{tn4LorRota123}
\end{equation}%
where $\eta$ is independent of $\alpha_1,\alpha_2$ and $\alpha_3$.

To solve equation (\ref{tn4dpsi}), we note that the equation can be written as
\begin{equation}
\partial _{\psi }\eta +\left[f(r,\theta)(\Gamma ^{{\theta}{\phi}}+\Gamma ^{{\psi}{r}})+g(r,\theta)(
\Gamma ^{{r}{\phi}}-\Gamma ^{{\psi}{\theta}})\right]\eta=0
\label{tn4dpsia}
\end{equation}
where
\begin{eqnarray}
f(r,\theta)&=&\frac{(r^2+a^2+2ar\cos\theta)^{3/2}n_1+an_2r^2\cos\theta+n_2r^3}
{4(r^2+a^2+2ar\cos\theta)^{1/2}\{(r^2+a^2+2ar\cos\theta)^{1/2}(r+n_1)+n_2r\}^2
}\nonumber\\
&&\\
g(r,\theta)&=&\frac{an_2r^2\sin\theta}{4(r^2+a^2+2ar\cos\theta)^{1/2}\{(r^2+a^2+2ar\cos\theta)^{1/2}(r+
n_1)+n_2r\}^2
}\nonumber\\
&&
\end{eqnarray}
So, the solution to equation (\ref{tn4dpsia}) satisfies 
\begin{equation}
\Gamma ^{{\psi}{r}{\theta}{\phi}}\eta =\eta 
\label{projprthph}
\end{equation}%
This equation eliminates another half of the supersymmetry provided $\eta$ is
independent of $\psi$, too. With this projection operator, (\ref{tn4dtheta}) and
(\ref{tn4dphi}) can be solved to give 
\begin{equation}
\eta =\exp \left\{ -\frac{\theta }{2}\Gamma ^{\hat{\psi}\hat{\phi}%
}\right\} \exp \left\{ \frac{\phi }{2}\Gamma ^{\hat{\theta}\hat{\phi}%
}\right\}\lambda  \label{tn4epsilon}
\end{equation}
where $\lambda$ is independent of $\theta$ and $\phi$.
Finally, we conclude due to two projections (\ref{proj012}) and (\ref{projprthph}), embedding Gibbons-Hawking space in M2 metric preserves 1/4
of supersymmetry.

Next, we consider the M5-brane solutions considered in section 5, given by (\ref{M5sol1}). Substituting $\varepsilon=H^{-1/12}\epsilon$ in the Killing spinor equations 
(\ref {killingspinoreq2}) yields
\begin{equation}
\Gamma^{tx_1x_2x_3x_4x_5}\epsilon=\epsilon\label{pro5}
\end{equation}
We note that for the anti-M5-brane $\alpha=-1$ in (\ref{FScompo}), the projection equation (\ref{pro5}) changes to $\Gamma^{tx_1x_2x_3x_4x_5}\epsilon=-\epsilon$. Moreover, we get three equations for $\epsilon$ that are given exactly by equations (\ref{tn4dpsi}), (\ref{tn4dtheta}) and (\ref{tn4dphi}). 
The solutions to these three equations imply
\begin{equation}
\Gamma ^{{\psi}{r}{\theta}{\phi}}\epsilon=\epsilon \label{proM5}
\end{equation}
and
\begin{equation}
\epsilon =\exp \left\{ -\frac{\theta }{2}\Gamma ^{\hat{\psi}\hat{\phi}%
}\right\} \exp \left\{ \frac{\phi }{2}\Gamma ^{\hat{\theta}\hat{\phi}%
}\right\}\xi  \label{tn4epsilonM5}
\end{equation}
where $\xi$ is independent of $\theta$ and $\phi$.

So, the two projection operators given by (\ref{pro5}) and (\ref{proM5}) show M5-brane solutions preserve 1/4 of supersymmetry.

Finally we consider how much supersymmetry could be preserved by the solutions (\ref{tn4xtn4mtrc}) with metric function (\ref{HTN4TN4}), given in section \ref{sec:M2mix}. 

As in the case of M2-brane, we get the projection equation
\begin{equation}
\Gamma ^{{t}{x}_{1}{x}_{2}}\epsilon =-\epsilon\label{p0}
\end{equation}
that remove half the supersymmetry, 
after substituting
$\varepsilon=H^{-1/6}\epsilon$ into the Killing spinor equations 
(\ref {killingspinoreq2}). The remaining equations could be solved by considering
\begin{eqnarray}
\Gamma^{\psi r\theta\phi}\epsilon&=&\epsilon\label{p1}\\
\Gamma^{\alpha_3y\alpha_1\alpha_2}\epsilon&=&\epsilon\label{p2}
\end{eqnarray}
However, the three projection operators in (\ref{p0}),(\ref{p1}) and (\ref{p2}) are not independent, since their indices altogether cover all the non-coordinate tangent space. Hence, we have only two independent projection operators, meaning 1/4 of the
supersymmetry is preserved.
 
\section{Decoupling Limits of Solutions}

\label{sec:dec}

In this section we consider the decoupling limits for the various
solutions we have presented above. The specifics of calculating the
decoupling limit are shown in detail elsewhere (see for example \cite 
{DecouplingLim}), so we will only provide a brief outline here. The process
is the same for all cases, so we will also only provide specific examples of
a few of the solutions above.

At low energies, the dynamics of the D2 brane decouple from the bulk, with
the region close to the D6 brane corresponding to a range of energy scales
governed by the IR fixed point \cite{DecouplingLim1}. For D2 branes localized
on D6 branes, this corresponds in the field theory to a vanishing mass for
the fundamental hyper-multiplets. Near the D2 brane horizon ($H\gg 1$), the
field theory limit is given by 
\begin{equation}
g_{YM2}^{2}=g_{s}\ell _{s}^{-1}=\text{fixed.}  \label{gymFTlimit}
\end{equation} 
In this limit the gauge couplings in the bulk go to zero, so the dynamics
decouple there. In each of our cases above, we scale the coordinates $y$ and $r$
such that 
\begin{equation}
Y=\frac{y}{\ell _{s}^{2}},~~~~U=\frac{r}{\ell _{s}^{2}}
\label{YUdecoupling1}
\end{equation} 
are fixed (where $Y$ and $U$, are used where appropriate). As an
example we note that this will change the harmonic function of the D6 brane
in the Gibbons-Hawking case to the following (recall that to avoid any conical singularity, we should have $n_1=n_2=n$, hence the
asymptotic radius of the 11th dimension is $R_{\infty }=n=g_{s}\ell _{s}
$) 
\begin{equation}
V_\epsilon(U,\theta)=\epsilon+g_{YM2}^{2}N_{6}\{\frac{1}{U}+\frac{1}{\sqrt{U^2+A^2+2AU\cos\theta}}\}
\label{F2}
\end{equation} 
where we rescale $a$ to $a=A\ell _{s}^{2}$ and generalize to the case of $N_{6}$ D6 branes. 
We notice that the metric function $H(y,r,\theta)$ scales as $H(Y,U,\theta)=\ell
_{s}^{-4}h(Y,U,\theta)$ if the coefficients $f_{c,M},f'_{c,M},\cdots$ obey some specific scaling. The scaling behavior of $H(Y,U,\theta)$ causes then the D2-brane to warp the ALE region and the asymptotically flat region of the D6-brane geometry. As an example, we calculate $h(Y,U,\theta)$ that corresponds to (\ref{2nutmetric}). It is given by 
\begin{eqnarray}
h(Y,U,\theta) &=&32\pi^2N_2g_{YM}^2\int_0^\infty dC \int_0^\infty d{\cal M}\, \frac{J_1(CY)}{Y}\times\nonumber\\
&\times&
\left\{
\tilde{\mathcal{H}}_{C}(\Omega,g_{YM})\{F_{C,{\cal M}}+F_{C,{\cal M}}^{\prime}\ln\left\vert
1-\frac{\Omega}{A}\right\vert \}\delta_{A,\Omega_0}+F_{C,{\cal M}}^{\prime}\sum_{n=0}^{\infty}b_{n,\Omega_0} 
(1-\frac{\Omega}{\Omega_0})^{n}\right\}
\times \nonumber \\
&\times& 
\left\{  \tilde{\mathcal{H}}_{C}(\Lambda,g_{YM})\{G_{C,{\cal M}}+G_{C,{\cal M}} 
^{\prime}\ln\left\vert 1-\frac{\Lambda}{A}\right\vert \}\delta_{A,\Lambda_0}+G_{C,{\cal M}}^{\prime
}\sum_{n=0}^{\infty}d_{n,\Lambda_0}(1-\frac{\Lambda}{\Lambda_0})^{n}\right\}.\nonumber\\
&&
\label{hTN2}
\end{eqnarray} 
where we rescale $c=C/\ell _{s}^{2}$ and $M={\cal M}\ell _{s}^{4}$. We notice that decoupling demands rescaling of the coefficients $f_{c,M},f'_{c,M},\cdots$ in (\ref{2nutmetric}) by
$f_{c,M}=F_{C,{\cal M}}/\ell _s^6,f'_{c,M}=F'_{C,{\cal M}}/\ell _s^6,\cdots$. In (\ref{hTN2}), $\Omega=\sqrt{U^2+A^2+2AU\cos\theta}+U$ and $\Lambda=\sqrt{U^2+A^2+2AU\cos\theta}-U$ and we use $\ell _{p}=g_{s}^{1/3}\ell _{s}$ to
rewrite $Q_{M2}=32\pi ^{2}N_{2}\ell _{p}^{6}$ in terms of $\ell_s$ given by $Q_{M2}=32\pi ^{2}N_{2}g_{YM2}^{4}\ell _{s}^{8}$.

The respective ten-dimensional supersymmetric metric (\ref{ds10TN4}) scales as
\begin{eqnarray}
ds_{10}^{2} &=&\ell _{s}^{2}\{h^{-1/2}(Y,U,\theta)V_\epsilon^{-1/2}(U,\theta)\left(
-dt^{2}+dx_{1}^{2}+dx_{2}^{2}\right) +  \nonumber \\
&+&h^{1/2}(Y,U,\theta)V_\epsilon^{-1/2}(U,\theta)\left( dY^{2}+Y^{2}d\Omega _{3}^{2}\right)+\nonumber\\
&+&h^{1/2}(Y,U,\theta)V_\epsilon^{1/2}(U,\theta)(dU^{2}+U^{2}d\Omega _{2}^{2})
\}
\label{m10}
\end{eqnarray} 
and so there is only one overall normalization factor of $\ell _{s}^{2}$ in the
metric (\ref{m10}). This is the expected result for a solution that is a
supergravity dual of a QFT. The other M2-brane and supersymmetric ten-dimensional solutions, given by  (\ref{secsolM2}), (\ref{2nutmetric}), (\ref{HTN4TN4}) and (\ref{ds10tnXtn}) have qualitatively the same behaviors in decoupling limit.

We now consider an analysis of the decoupling limits of M5-brane solution given by metric function (\ref{M5sol1}). 

At low energies, the dynamics of IIA NS5-branes will decouple from the
bulk \cite{DecouplingLim2}. Near the NS5-brane horizon ($H>>1$), we are interested in the
behavior of the NS5-branes in the limit where string coupling vanishes 
\begin{equation}
g_{s}\rightarrow 0  \label{gym1}
\end{equation} 
and 
\begin{equation}
\ell _{s}=\text{ fixed.}  \label{gym2}
\end{equation} 
In these limits, we rescale the radial coordinates such that they can be
kept fixed 
\begin{equation}
Y=\frac{y}{g_{s}\ell _{s}^{2}},~U=\frac{r}{g_{s}\ell _{s}^{2}}.
\label{yrrescale}
\end{equation} 
This causes the harmonic function of the D6-brane for the Gibbons-Hawking solution (\ref{tng10}), change to 
\begin{equation}
V_\epsilon(r,\theta)=\epsilon+\frac{N_6}{\ell _{s}}\{\frac{1}{U}+\frac{1}{\sqrt{U^2+A^2+2AU\cos\theta}}\}\equiv V_\epsilon (U,\theta)
\end{equation} 
where we generalize to $N_{6}$ D6-branes and rescale $a=A\ell _{s}^{2}g_{s}$.

We can show the harmonic function for the NS5-branes (\ref{M5sol1}) rescales according to $H(Y,U,\theta)=g_{s}^{-2}h(Y,U,\theta)$. In fact, we have

\begin{eqnarray}
H(Y,U,\theta) &=&
\frac{\pi N_5\ell_s^5}{g_s^2}\int_0^\infty dC \int_0^\infty d{\cal M} \cos(CY+\zeta)\times\nonumber\\
&\times&\left\{
\tilde{\mathcal{H}}_{C}(\Omega,\ell_s)\{F_{C,{\cal M}}+F_{C,{\cal M}}^{\prime}\ln\left\vert
1-\frac{\Omega}{A}\right\vert \}\delta_{A,\Omega_0}+F_{C,{\cal M}}^{\prime}\sum_{n=0}^{\infty}b_{n,\Omega_0} 
(1-\frac{\Omega}{\Omega_0})^{n}\right\}
\times \nonumber \\
&\times& 
\left\{  \tilde{\mathcal{H}}_{C}(\Lambda,\ell_s)\{G_{C,{\cal M}}+G_{C,{\cal M}} 
^{\prime}\ln\left\vert 1-\frac{\Lambda}{A}\right\vert \}\delta_{A,\Lambda_0}+G_{C,{\cal M}}^{\prime
}\sum_{n=0}^{\infty}d_{n,\Lambda_0}(1-\frac{\Lambda}{\Lambda_0})^{n}\right\}.\nonumber\\
&&
\label{M5sol11}
\end{eqnarray} 
where we use $\ell _{p}=g_{s}^{1/3}\ell _{s}$ to rewrite 
$Q_{M5}=\pi N_{5}\ell _{p}^{3}$ as $\pi N_{5}g_{s}\ell _{s}^{3}$.
To get (\ref{M5sol11}), we rescale $c=C/(g_{s}\ell _{s}^{2})$, $M={\cal M}g_s^2\ell_s^4$ and $a=Ag_s\ell_s^2$ such that $h(Y,U,\theta)$ doesn't have any $g_{s}$ dependence.

In decoupling limit, the ten-dimensional metric (\ref{tng10}) becomes,
\begin{eqnarray}
ds_{10}^{2} &=&V_\epsilon^{-1/2}(U,\theta)\left(
-dt^{2}+dx_{1}^{2}+dx_{2}^{2}+dx_{3}^{2}+dx_{4}^{2}+dx_{5}^{2}\right) +\ell_s^4\{h(Y,U,\theta) 
V_\epsilon^{-1/2}(U,\theta)dY^{2}+  \nonumber \\
&+&h(Y,U,\theta)V_\epsilon^{1/2}(U,\theta)\left( dU^{2}+U^{2}d\Omega _{2}^{2}\right)\}.
\label{tng100}
\end{eqnarray}

In the limit of vanishing $g_{s}$\ with fixed $l_{s}$\ (as we did in 
(\ref{gym1}) and (\ref{gym2})), the decoupled free theory on NS5-branes should be
a little string theory \cite{shiraz} (i.e. a 6-dimensional non-gravitational
theory in which modes on the 5-brane interact amongst themselves, decoupled
from the bulk). We note that our NS5/D6 system is obtained from M5-branes by
compactification on a circle of self-dual transverse geometry. Hence the IIA
solution has T-duality with respect to this circle. The little string theory
inherits the same T-duality from IIA string theory, since taking the limit
of vanishing string coupling commutes with T-duality. Moreover T-duality
exists even for toroidally compactified little string theory. In this case,
the duality is given by an $O(d,d,\mathbb{Z})$ symmetry where $d$\ \ is the
dimension of the compactified toroid. These are indications that the little
string theory is non-local at the energy scale $l_{s}^{-1}$ and in
particular in the compactified theory, the energy-momentum tensor can't be
defined uniquely \cite{aha}.

As the last case, we consider the analysis of the decoupling limits of the IIB solution
that can be obtained by T-dualizing the compactified M5-brane solution (\ref{ds11m5p}). 
The type IIA NS5$\perp $ 
D6(5) configuration is given by the metric (\ref{tng10}) and fields (\ref{dilTN}), ( 
\ref{tnH3}), (\ref{tnRR}) and (\ref{tnA3}).

We apply the T-duality \cite{Cascaless} in the $x_{1}-$direction of the metric ( 
\ref{tng10}), that yields 
 gives the IIB dilaton field 
 \begin{equation}
 \widetilde{\Phi }=\frac{1}{2}\ln \frac{H}{\tilde{f}}  \label{IBdilaton}
 \end{equation} 
the 10D type IIB metric, as 
\begin{eqnarray}
\widehat{ds}_{10}^{2} &=&V_\epsilon^{-1/2}(r,\theta)\left(
-dt^{2}+V_\epsilon(r,\theta)dx_{1}^{2}+dx_{2}^{2}+dx_{3}^{2}+dx_{4}^{2}+dx_{5}^{2}\right) +\nonumber \\
&+&H(y,r,\theta) 
V_\epsilon^{-1/2}(r,\theta)dy^{2}+  
H(y,r,\theta)V_\epsilon^{1/2}(r,\theta)\left( dr^{2}+r^{2}d\Omega _{2}^{2}\right).
\label{IIBmetric}
\end{eqnarray} 
The metric (\ref{IIBmetric}) describes a IIB NS5$\perp $D5(4) brane
configuration (along with the dualized dilaton, NSNS and RR fields).

At low energies, the dynamics of IIB
NS5-branes will decouple from the bulk. Near the NS5-brane horizon ($H>>1$),
the field theory limit is given by 
\begin{equation}
g_{YM5}=\ell _{s}=\text{ fixed}  \label{gym}
\end{equation} 
We rescale the radial coordinates $y\ $and $r$\ as in (\ref{yrrescale}),
such that their corresponding rescaled coordinates $Y$\ and $U$ are kept
fixed. The harmonic function of the D5-brane is 
\begin{equation}
V_\epsilon(r,\theta)=\epsilon+\frac{N_5}{g_{YM5}}\{\frac{1}{U}+\frac{1}{\sqrt{U^2+A^2+2AU\cos\theta}}\}
\end{equation} 
where ${N}_{5}$\ is the number of D5-branes.

The harmonic function of the NS5$\perp $D5 system (\ref{IIBmetric}), 
rescales according to $ 
H(Y,U,\theta)=g_{s}^{-2}{h}(Y,U,\theta)$, where 
\begin{eqnarray}
{h}(Y,U,\theta)&=&\pi N_5 g_{YM5}^5\int_0^\infty dC \int_0^\infty d{\cal M} \cos(CY+\zeta)\times\nonumber\\
&\times&\left\{
\tilde{\mathcal{H}}_{C}(\mu,g_{YM5})\{F_{C,{\cal M}}+F_{C,{\cal M}}^{\prime}\ln\left\vert
1-\frac{\Omega}{A}\right\vert \}\delta_{A,\Omega_0}+F_{C,{\cal M}}^{\prime}\sum_{n=0}^{\infty}b_{n,\Omega_0} 
(1-\frac{\Omega}{\Omega_0})^{n}\right\}
\times \nonumber \\
&\times& 
\left\{  \tilde{\mathcal{H}}_{C}(\lambda,g_{YM5})\{G_{C,{\cal M}}G_{C,{\cal M}} 
^{\prime}\ln\left\vert 1-\frac{\Lambda}{A}\right\vert \}\delta_{A,\Lambda_0}+G_{C,{\cal M}}^{\prime
}\sum_{n=0}^{\infty}d_{n,\Lambda_0}(1-\frac{\Lambda}{\Lambda_0})^{n}\right\}.\nonumber\\
&&
\end{eqnarray} 
In this case, the ten-dimensional metric (\ref{IIBmetric}), in the
decoupling limit, becomes 
\begin{eqnarray}
\widetilde{ds}_{10}^{2} &=&V_\epsilon^{-1/2}(U,\theta)\left(
-dt^{2}+V_\epsilon(U,\theta)dx_{1}^{2}+dx_{2}^{2}+dx_{3}^{2}+dx_{4}^{2}+dx_{5}^{2}\right)+
 \nonumber \\
&+&g_{YM5}^2h(Y,U,\theta) 
\{V_\epsilon^{-1/2}(U,\theta)dY^{2}+ 
+V_\epsilon^{1/2}(U,\theta)\left( dU^{2}+U^{2}d\Omega _{2}^{2}\right)\}.
\end{eqnarray}

The decoupling limit illustrates that the decoupled theory in the low energy
limit is super Yang-Mills theory with $g_{YM}=\ell _{s}.$\ In the limit of
vanishing $g_{s}$\ with fixed $l_{s}$,\ the decoupled free theory on IIB
NS5-branes (which is equivalent to the limit $g_{s}\rightarrow \infty $\ of
decoupled S-dual of the IIB D5-branes) reduces to a IIB (1,1) little string
theory with eight supersymmetries.\

\section{Concluding Remarks}

The central thrust of this paper is the explicit and exact construction of supergravity
solutions for fully localized D2/D6 and NS5/D6 brane intersections without restricting
to the near core region of the D6 branes. Unlike all the other known solutions, the novel feature of these solutions is the dependence of the metric function to three (and four) transverse coordinates. 
These exact solutions are new M2 and M5 brane metrics that
are presented in equations (\ref{a32}), (\ref{a322}), (\ref
{secsolM2}), (\ref{2nutmetric}), (\ref{M5sol1}) and (\ref{HTN4TN4}) which are the main results of this paper.
The common feature of all of these solutions is that the brane function is a
convolution of an decaying function with a damped
oscillating one. The metric functions vanish far from the M2 and M5 branes and
diverge near the brane cores.

Dimensional reduction of the M2 solutions to ten dimensions gives us
intersecting IIA D2/D6 configurations that preserve 1/4 of the
supersymmetry. For the M5 solutions, dimensional reduction yields IIA
NS5/D6 brane systems overlapping in five directions. The latter solutions also preserve 1/4 of the supersymmetry 
and in both cases the reduction yields metrics with acceptable
asymptotic behaviors. 

We considered the decoupling limit of our solutions and found that D2 and NS5 branes
can decouple from the bulk, upon imposing proper scaling on some of the coefficients in the integrands.

In the case of M2 brane solutions; when the D2 brane decouples from the bulk, the theory on
the brane is 3 dimensional $\mathcal{N}=4$ $SU($N$_{2})$ super Yang-Mills
(with eight supersymmetries) coupled to N$_{6}$\ massless hypermultiplets  
\cite{pelc}. This point is obtained from dual field theory and since our solutions preserve the same amount of supersymmetry, a similar dual field description should be attainable. 

In the case of M5 brane solutions; the resulting theory on the NS5-brane in the
limit of vanishing string coupling with fixed string length is a little
string theory. In the standard case, the system of N$_{5}$\ NS5-branes located at N$_{6}$\
D6-branes can be obtained by dimensional reduction of \ N$_{5}$N$_{6}$\
coinciding images of M5-branes in the flat transverse geometry. In this
case, the world-volume theory (the little string theory) of the IIA
NS5-branes, in the absence of D6-branes, is a non-local non-gravitational
six dimensional theory \cite{seiberg}. This theory has (2,0) supersymmetry
(four supercharges in the \textbf{4}\ representation of Lorentz symmetry $ 
Spin(5,1)$) and an R-symmetry $Spin(4)$ remnant of the original ten
dimensional Lorentz symmetry. The presence of the D6-branes breaks the
supersymmetry down to (1,0), with eight supersymmetries. Since we found that
some of our solutions preserve 1/4 of supersymmetry, we expect that the
theory on NS5-branes is a new little string theory. \ 
By T-dualization of the 10D IIA theory along a direction parallel to the
world-volume of the IIA NS5, we find a IIB NS5$\perp $D5(4) system,
overlapping in four directions. The world-volume theory of the IIB
NS5-branes, in the absence of the D5-branes, is a little string theory with
(1,1) supersymmetry. The presence of the D5-brane, which has one transverse
direction relative to NS5 world-volume, breaks the supersymmetry down to
eight supersymmetries. This is in good agreement with the number of
supersymmetries in 10D IIB theory: T-duality preserves the number of
original IIA supersymmetries, which is eight. Moreover we conclude that the
new IIA and IIB little string theories are T-dual: the actual six
dimensional T-duality is the remnant of the original 10D T-duality after
toroidal compactification.

A useful application of the exact M-brane solutions in our paper is to
employ them as supergravity duals of the NS5 world-volume theories with
matter coming from the extra branes. More specifically, these solutions can
be used to compute some correlation functions and spectrum of fields of our
new little string theories.

In the standard case of $A_{k-1}$ (2,0) little string theory, there is an
eleven dimensional holographic dual space obtained by taking appropriate
small $g_s$ limit of an M-theory background corresponding to M5-branes with
a transverse circle and $k$ units of 4-form flux on $S^3 \otimes S^1$. In
this case, the supergravity approximation is valid for the (2,0) little
string theories at large $k$ and at energies well below the string scale.
The two point function of the energy-momentum tensor of the little string
theory can be computed from classical action of the supergravity evaluated
on the classical field solutions \cite{shiraz}.

Near the boundary of the above mentioned M-theory background, the string
coupling goes to zero and the curvatures are small. Hence it is possible to
compute the spectrum of fields exactly. In \cite{aha}, the full spectrum
of chiral fields in the little string theories was computed and the results
are exactly the same as the spectrum of the chiral fields in the low energy
limit of the little string theories. Moreover, the holographic dual theories
can be used for computation of some of the states in our little string
theories.

We conclude with a few comments about possible directions for future work.
Investigation of the different regions of the metric (\ref{ds11m5p}) or
alternatively the 10D string frame metric (\ref{tng100}) with a
dilaton (also for other considered EH and TB cases) for small and large
Higgs expectation value $U$\ would be interesting, as it could provide a
means\ for finding a holographical dual relation to the new little string
theory we obtained. Moreover, the Penrose limit of the near-horizon geometry
may be useful for extracting information about the high energy spectrum of
the dual little string theory \cite{Gomiss}. The other open issue is the
possibility of the construction of a pp-wave spacetime which interpolates
between the different regions of the our new IIA NS5-branes. Moreover, it would be interesting 
(and of course very complicated) to find the exact analytic solutions for the brane functions with the embedded Gibbons-Hawking spaces with $k>2$.

\vspace*{1cm}

{\Large Acknowledgments} 

\vspace*{0.5cm}

This work was supported by the Natural Sciences and
Engineering Research Council of Canada.

\vspace*{1cm}

\appendix

{\section {The Heun-C functions}}

\vspace*{0.5cm}

The Heun-C function $\mathcal{H}_{C}(\alpha,\beta,\gamma,\delta,\lambda,z)$ 
is the solution to the confluent Heun's differential equation \cite{Heuref}
\begin{equation}
\mathcal{H}_{C}''+(\alpha+\frac{\beta+1}{z}+\frac{\gamma+1}{z-1})\mathcal{H}_{C}'+
(\frac{\mu}{z}+\frac{\nu}{z-1})\mathcal{H}_{C}=0
\label{Heq}
\end{equation}
where $\mu=\frac{\alpha-\beta-\gamma+\alpha\beta-\beta\gamma}{2}-\lambda$ and
$\nu=\frac{\alpha+\beta+\gamma+\alpha\beta+\beta\gamma}{2}+\delta+\lambda$.
The equation (\ref{Heq}) has two regular singular points at $z=0$ and $z=1$ and 
one irregular singularity at $z=\infty$. The $\mathcal{H}_{C}$ function is 
regular around the regular singular point $z=0$ and is given by
$\mathcal{H}_{C}=\Sigma_{n=0}^\infty h_n(\alpha,\beta,\gamma,\delta,\lambda)z^n$,
where $h_0=1$. The series is convergent on the unit disk $\vert z\vert <1$ and the coefficients $h_n$ are determined by the recurrence relation
\begin{equation}
h_n=\Theta_nh_{n-1}+\Phi_nh_{n-2}
\end{equation} 
where we set $h_{-1}=0$ and 
\begin{eqnarray}
\Theta_n&=&\frac{2n(n-1)+(1-2n)(\alpha-\beta-\gamma)+2\lambda-\alpha\beta+\beta\gamma}{2n(n+\beta)}     \\
\Phi_n&=&\frac{\alpha(\beta+\gamma+2(n-1))+2\delta}{2n(n+\beta)}.
\end{eqnarray}

\vspace{0.25cm}

{\section {Coefficients of Series in (\ref{a32}) }}

\vspace*{0.5cm}

Here we list some coefficients that appear in (\ref{a32})
\begin{eqnarray}
b_{0,\mu_0 > a}&=&1\nonumber\\
b_{1,\mu_0 > a}&=&-\mu_0\nonumber\\
b_{2,\mu_0 > a}&=&\{-\frac{\mu _{0}}{(\mu _{0}^{2}-a^{2})}+\frac{c^{2}(\epsilon \mu _{0}^{2}+4M+2N_{+}\mu
_{0})}{8(\mu _{0}^{2}-a^{2})}\}\mu_0^2\nonumber\\
b_{3,\mu_0 > a}&=&\{\frac{c^{2}(\epsilon \mu
_{0}^{3}+8\mu _{0}M+3N_+\mu _{0}^{2}+N_{+}a^{2}+\epsilon \mu _{0}a^{2})}{ 
12(\mu _{0}^{2}-a^{2})^{2}}\nonumber\\
&+&\frac{-24\mu _{0}^{2}-8a^{2}-c^{2}\epsilon \mu
_{0}^{4}+c^{2}\epsilon \mu _{0}^{2}a^{2}-4c^{2}M\mu
_{0}^{2}+4c^{2}Ma^{2}-2c^{2}N_+\mu _{0}^{3}+2c^{2}N_{+}\mu _{0}a^{2}}{24(\mu
_{0}^{2}-a^{2})^{2}}\}\mu_0^3 \nonumber\\
&&\label{BBS}\\
d_{0,\vert\lambda_0\vert < a}&=&1\nonumber\\
d_{1,\vert\lambda_0\vert < a}&=&-\lambda_0\nonumber\\
d_{2,\vert\lambda_0\vert < a}&=&\{-\frac{c^{2}(\epsilon \lambda_0^2+4M+2N_-\lambda_0)}{8(a^{2}-\lambda _{0}^{2})}+\frac{\lambda _{0}}{(a^{2}-\lambda
_{0}^{2})}\}\lambda_0^2\nonumber\\
d_{3,\vert\lambda_0\vert < a}&=&\{\frac{c^{2}(\epsilon \lambda _{0}^{3}+8\lambda
_{0}M+3N_{-}\lambda _{0}^{2}+N_{-}a^{2}+\epsilon \lambda _{0}a^{2})}{ 
12(\lambda _{0}^{2}-a^{2})^{2}}+\nonumber\\
&+&\frac{-24\lambda
_{0}^{2}-8a^{2}-c^{2}\epsilon \lambda _{0}^{4}+c^{2}\epsilon \lambda
_{0}^{2}a^{2}-4c^{2}M\lambda _{0}^{2}+4c^{2}Ma^{2}-2c^{2}N_{-}\lambda
_{0}^{3}+2c^{2}N_{-}\lambda _{0}a^{2}}{24(a^{2}-\lambda _{0}^{2})^{2}} 
\}\lambda_0^3.\nonumber\\
&&\label{DDS}
\end{eqnarray}

The recursion relations that we have used to derive the coefficients (\ref{BBS}) and (\ref{DDS}), both are in the form of
\begin{equation}
Q_n={\cal Q}_1Q_{n-1}+{\cal Q}_2Q_{n-2}+{\cal Q}_3Q_{n-3}+{\cal Q}_4Q_{n-4}\label{REC}
\end{equation} 
where $n\geq 2$ and $Q_0=Q_1=1$. Moreover $Q_{n<0}=0$. The coefficients (\ref{BBS}) are related to $Q$'s by 
\begin{equation}
b_{n,\mu_0>a}=(-\mu_0)^nQ_n
\end{equation}
and the functions ${\cal Q}$ depend on $\epsilon,\mu_0,n,c,a,M,N_+$. For (\ref{DDS}), the relation to $Q$'s is
\begin{equation}
d_{n,\vert \lambda_0 \vert <a}=(-\lambda_0)^nQ_n
\end{equation}
where the functions ${\cal Q}$ depend on $\epsilon,\mu_0,n,c,a,M,N_-$. In both cases, the radius of convergence is large enough to find the membrane function (\ref{2nutmetric}) at many intermediate-zone points. As an example,
for the choice of $a=\epsilon=M=1,c=N_+=2$ and $\mu_0=10.75$, the series is divergent for
$0.9906 < \mu < 20.5093$

\vspace{0.25cm}

{\section {Representation of Clifford Algebra}}

\vspace*{0.5cm}

The gamma matrices satisfy the Clifford Algebra 
\begin{equation}
\left\{ \Gamma _{a},\Gamma _{b}\right\} =-2\eta _{ab}  \label{CliffordAlg}
\end{equation}%
where we are using the Lorentzian signature $\left[ -1,+1,\ldots ,+1\right] $%
. A representation of the algebra (\ref{CliffordAlg}) is given by 
\begin{equation}
\Gamma _{\xi}=\gamma _{\xi}\otimes 1  \label{GAmmaihat}
\end{equation}%
and 
\begin{equation}
\Gamma _{\Xi +4}=\gamma _{5}\otimes \widehat{\Gamma }_{\Xi }
\end{equation}%
where $\xi=0,1,2,3$ and $\Xi =0,1,...,6$\ denotes the spacetime
indices for the tangent space groups $SO(1,3)$\ and $SO(7)$. The $\Gamma
_{\Xi +4}$\ (and $\widehat{\Gamma }_{\Xi }$) satisfy the anticommutation
relations 
\begin{equation}
\{\Gamma _{\Xi +4}^{\text{ \ \ }},\Gamma _{\Psi +4}^{\text{ \ \ }}\}=\{%
\widehat{\Gamma }_{\Xi }^{\text{ \ \ }},\widehat{\Gamma }_{\Psi }^{\text{ \
\ }}\}=-2\delta _{\Xi \Psi}  \label{Cliff8}
\end{equation}%
where the $\widehat{\Gamma }_{\Xi }$'s are given by{\large \ } 
\begin{equation}
\begin{array}{c}
\widehat{\Gamma }_{0}=i\gamma _{0}\otimes 1 \\ 
\widehat{\Gamma }_{i}=\gamma _{i}\otimes 1 \\ 
\widehat{\Gamma }_{i+3}=i\gamma _{5}\otimes \sigma _{i}%
\end{array}
\label{Gammahats}
\end{equation}%
in terms of the Pauli matrices $\sigma _{i}$ $(i=1,2,3)$, $\gamma
_{0}=\left( 
\begin{array}{cc}
0 & 1 \\ 
1 & 0%
\end{array}%
\right) $, and $\gamma _{5}=i\gamma _{0}\gamma _{1}\gamma _{2}\gamma _{3}$.


\vspace*{1cm}

\begin{thebibliography}{99}                                                                                               

\bibitem {gr1}E. Witten, \textit{Nucl. Phys.} \textbf{B443} (1995) 85.

\bibitem {gr2}M.J. Duff, J.T. Liu and R. Minasian, \textit{Nucl. Phys.}
\textbf{B452} (1995) 261.

\bibitem {gr3}J.H. Schwarz, \textit{Phys. Lett.} \textbf{B367} (1996) 97.

\bibitem {Tsey}A.A. Tseytlin, \textit{Nucl. Phys.} \textbf{B475} (1996) 149.

\bibitem {oth}A. Loewy, \textit{\ Phys. Lett.} \textbf{B463} (1999) 41.

\bibitem {hashi}S.A. Cherkis and A. Hashimoto, \textit{JHEP}
\textbf{0211} (2002) 036.

\bibitem {CGMM2}R. Clarkson, A.M. Ghezelbash and R.B. Mann, \textit{JHEP}
\textbf{0404} (2004) 063; \textbf{0408} (2004) 025.

\bibitem {ATM2}A.M. Ghezelbash and R.B. Mann, \textit{JHEP} \textbf{0410}
(2004) 012.

\bibitem {GHEbianchiMbranes}A.M. Ghezelbash, \textit{Phys. Rev.} \textbf{D74}
(2006) 126004.

\bibitem {GH2resolvedconifolds}A.M. Ghezelbash, \textit{Phys. Rev.}
\textbf{D77} (2008) 026006.

\bibitem {IT}N. Itzaki, A.A. Tseytlin and S. Yankielowicz, 
\textit{Phys. Lett.} \textbf{B432} (1998) 298.

\bibitem{IT2} A. Hashimoto, \textit{JHEP} \textbf{9901}
(1999) 018.

\bibitem {Ali}S. Arapoglu, N.S. Deger and A. Kaya, \textit{Phys. Lett.}
\textbf{B578} (2004) 203.

\bibitem {cremmer}E. Cremmer, B. Julia, H. Lu and C.N. Pope,
\textit{Nucl. Phys.} \textbf{B523} (1998) 73.

\bibitem {DuffKK}M.J. Duff, B.E.W. Nilsson and C.N. Pope, \textit{Phys. Rep.}
\textbf{130} (1986) 1.

\bibitem {smith}D.J. Smith, \textit{Class. Quant. Grav.} \textbf{20} (2003) R233.

\bibitem{EgHan}T. Eguchi and A.J. Hanson, \textit{Phys. Lett.} \textbf{B74} (1978) 249.

\bibitem{forbps}
A.A. Tseytlin, \textit{Class. Quant. Grav.} \textbf{14} (1997) 2085;
\textit{Nucl. Phys.} \textbf{B487} (1997) 141. 

\bibitem{ex}
I.R. Klebanov and A.A. Tseytlin, \textit{Nucl. Phys.} \textbf{B475}
(1996) 179.

\bibitem{ex2}
J.P. Gauntlett, D.A. Kastor and J. Traschen, \textit{Nucl. Phys.} \textbf{B478}
(1996) 544.

\bibitem{gauntlettetal3} J.P. Gauntlett, J.B. Gutowski and S. Pakis, 
\textit{JHEP} \textbf{0312} (2003) 049.

\bibitem{FF}
J.M. Figueroa-O'Farrill, , \textit{Class. Quant. Grav.} \textbf{17} (2000) 2925.

\bibitem{DecouplingLim} J. Maldacena, \textit{Adv. Theor. Math. Phys.} \textbf{2} (1998) 231.

\bibitem{DecouplingLim1} J. Maldacena, \textit{Int. J. Theor. Phys.} 
\textbf{38} (1999) 1113.

\bibitem{DecouplingLim2} N. Itzhaki, J.M. Maldacena, J. Sonnenschein and S.
Yankielowicz, \textit{Phys. Rev.} \textbf{D58} (1998) 046004.

\bibitem{shiraz} S. Minwalla and N. Seiberg, \textit{JHEP} \textbf{9906} (1999) 007.

\bibitem{aha} O. Aharony, M. Berkooz, D. Kutasov and N. Seiberg, \textit{JHEP} 
\textbf{9810} (1998) 004.

\bibitem{Cascaless} J.F.G. Cascales and A.M. Uranga, \textit{JHEP} \textbf{0401} 
(2004) 021.

\bibitem{pelc} O. Pelc and R. Siebelink, \textit{Nucl. Phys.} \textbf{B558}
(1999) 127.

\bibitem{seiberg} N. Seiberg, \textit{Phys. Lett.} \textbf{B408} (1997) 98.

\bibitem{Gomiss} J. Gomis and H. Ooguri, \textit{Nucl. Phys.} \textbf{B635}
(2002) 106.

\bibitem{Heuref} P.P. Fiziev, arXiv:0904.0245

\end{thebibliography}
\end{document}